\documentclass[11pt,a4paper]{article}
\pdfoutput=1
\usepackage{jheppub}
\usepackage{rotating}
\usepackage{array}
\usepackage{relsize}     
\usepackage{amsmath}
\usepackage{slashed}
\usepackage{booktabs}
\usepackage[pdftex,table]{xcolor}
\usepackage{mathtools}
\usepackage{empheq}
\usepackage{float}
\usepackage{blindtext}

\newcommand{\y}{z}

\renewcommand{\d}{\text{d}}
\newcommand{\lint}{\mathlarger{\int}}

\newcommand{\eqsp}{\;}

\newcommand{\ldefine}{\vcentcolon=}

\newcommand{\sm}{\text{SM}}
\newcommand{\elm}{\text{EM}}
\newcommand{\elmnu}{\sm}

\newcommand{\p}{P}
\newcommand{\eq}[1]{#1}

\preprint{DESY 20-160, ULB-TH/20-15}

\title{Updated BBN constraints on electromagnetic decays of MeV-scale particles}

\author{Paul Frederik Depta$^\text{a}$,}
\author{Marco Hufnagel$^{\text{a}, \text{b}}$, and}
\author{Kai Schmidt-Hoberg$^\text{a}$}

\affiliation{$^\text{a}$ DESY, Notkestra\ss e 85, D-22607 Hamburg, Germany}
\affiliation{$^\text{b}$ Service de Physique Théorique, Université Libre de Bruxelles, Boulevard du Triomphe, CP225, B-1050 Brussels, Belgium}

\emailAdd{frederik.depta@desy.de}
\emailAdd{marco.hufnagel@desy.de}
\emailAdd{kai.schmidt-hoberg@desy.de}

\abstract{
In this work, we revise and update model-independent constraints from Big Bang Nucleosynthesis on MeV-scale particles $\phi$ which decay into photons and/or electron-positron pairs.
We use the latest determinations of primordial abundances and extend the analysis in \cite{Hufnagel:2018bjp} by including all spin-statistical factors
as well as inverse decays, significantly strengthening the resulting bounds in particular for small masses.
For a very suppressed initial abundance of $\phi$, these effects become ever more important and we find that even a pure `freeze-in' abundance can be significantly constrained.
In parallel to this article, we release the public code \texttt{ACROPOLIS}  which numerically solves the reaction network necessary to evaluate the effect of photodisintegration on the final light element abundances. As an interesting application, we re-evaluate a possible solution of the lithium problem due to the photodisintegration of beryllium and find that e.g.\ an ALP produced via freeze-in can lead to a viable solution.}

\keywords{}

\begin{document}
\maketitle

\section{Introduction}

Light and weakly coupled dark sectors (DSs) naturally appear in many extensions of the Standard Model (SM) and
can significantly impact the cosmological evolution of our Universe, potentially affecting predictions of primordial element abundances produced by Big Bang Nucleosynthesis (BBN) or observables related to the Cosmic Microwave Background (CMB). Given the remarkable overall agreement between the inferred abundances of light elements and the corresponding predictions within the SM, it is well known that any deviation from the standard cosmology for sub-MeV temperatures is strongly constrained~\cite{Shvartsman:1969mm,Steigman:1977kc,Scherrer:1987rr,Cyburt:2015mya,Masso:1995tw,Masso:1997ru,Cadamuro:2010cz, Cadamuro:2011fd,Millea:2015qra,Hufnagel:2017dgo,Hufnagel:2018bjp,Forestell:2018txr,Depta:2019lbe,Depta:2020wmr,Ghosh:2020vti,Kawasaki:2020qxm}. 
While early studies mainly concentrated on heavy decaying relics, there has been a strong interest in light DSs recently~\cite{Batell:2009di,Andreas:2012mt,Schmidt-Hoberg:2013hba,Essig:2013vha,Izaguirre:2013uxa,Batell:2014mga,Dolan:2014ska,Krnjaic:2015mbs,Dolan:2017osp,Izaguirre:2017bqb,Knapen:2017xzo,Beacham:2019nyx,Bondarenko:2019vrb,Filimonova:2019tuy}.
Particles with masses in the MeV-range in particular transition from the relativistic to the non-relativistic regime during BBN, requiring a full evaluation of the cosmological evolution in order to calculate the effects on BBN~\cite{Cadamuro:2011fd,Millea:2015qra,Hufnagel:2017dgo,Hufnagel:2018bjp,Forestell:2018txr,Depta:2019lbe,Depta:2020wmr,Kawasaki:2020qxm}.

In this work, we revise and update model-independent constraints from Big Bang Nucleosynthesis on MeV-scale particles which decay into photons and/or electron-positron pairs.
These decaying particles can affect the primordial element abundances in a number of ways:
{\it i)} the presence of any additional particles beyond those in the SM increases the overall energy density of the Universe which in turn directly impacts
the Hubble rate, implying that the proton-to-neutron ratio as well as the time-temperature relation are generally changed;
{\it ii)} decays into electromagnetic radiation will generally lead to entropy injection into the thermal bath of SM particles and therefore change the baryon-to-photon ratio $\eta$;
{\it iii)} late decays into photons or electrons may also photodisintegrate light nuclei formed in the first three minutes. A detailed analysis of these different effects has already been done in
 \cite{Hufnagel:2018bjp}. However, possible inverse decays of the decaying particle $\phi$ had not been taken into account in \cite{Hufnagel:2018bjp}, leading to conservative limits in general.
Nevertheless, inverse decays generically give a model-independent contribution to the abundance of $\phi$ and can be very important in some regions of parameter space.
Here we extend and revise the results obtained in \cite{Hufnagel:2018bjp} by
\begin{enumerate}
\item including all spin-statistical factors with full Bose-Einstein and Fermi-Dirac distribution functions,
\item including inverse decays for the decaying particle,
\item taking into account the correlation between $N_\text{eff}$ and $\eta$ in the Planck data, and
\item making use of the latest determinations of primordial abundances (and constraining $^3\text{He}/\text{D}$ instead of $^3\text{He}/\text{H}$).
\end{enumerate}
Overall we find that taking into account the inverse decays with full spin-statistical factors significantly strengthens the resulting bounds in particular for small masses of $\phi$. Also for the case that the initial abundance of $\phi$ particles is rather small as is the case for example if the DS temperature is smaller than the temperature in the SM, inverse decays will contribute to a `freeze-in' abundance of $\phi$ and again significantly strengthen the limits.

We also reconsider a possible new physics solution to the long-standing lithium problem of standard BBN which takes advantage of the different threshold energies of nuclei for photodisintegration~\cite{Kusakabe:2013sna,Poulin:2015woa,Salvati:2016jng,Kawasaki:2020qxm}.
To evaluate the effect of photodisintegration on the final light element abundances we employ the public code \texttt{ACROPOLIS}~\cite{Depta:2020mhj} which we release in parallel to this article.
We find that indeed regions of parameter space exist in simple models which are consistent with all inferred primordial abundances including lithium, for example vanilla type axion-like particles (ALPs) produced via freeze-in.

This article is structured as follows. In the next section we give details of the cosmological evolution of the decaying particle $\phi$, paying particular attention to the effects of inverse decays on the $\phi$ abundance. In section~\ref{sec:BBN} we describe the procedure employed for calculating BBN limits. Section~\ref{sec:results} contains our results assuming a DS temperature which is generally different
from the temperature in the SM sector as an initial condition. The case with initially vanishing dark sector temperature is equivalent to the case of pure freeze-in. In section~\ref{sec:li} we finally discuss possible solutions to the long-standing lithium problem before we conclude in section~\ref{sec:conclusions}.

\section{Evolution and influence of the dark sector}

\subsection{Evolution of the decaying particle $\phi$}
\label{sec:evo_phi}

We consider a bosonic DS particle $\phi$ with mass $m_\phi$ and lifetime $\tau_\phi$ that decouples chemically from the DS at time $t_\text{cd}$ and corresponding DS temperature $T_\text{D,cd} = \zeta_\text{cd} T_\text{cd}$, where $T_\text{cd}$ is the SM temperature at $t_\text{cd}$.\footnote{Note that a longer phase of kinetic equilibrium after chemical decoupling does not impact our results~\cite{Hufnagel:2017dgo}.}
At chemical decoupling, its phase-space distribution function at momentum $p$ is given by the Bose-Einstein distribution,
\begin{align}
f_\phi(t_\text{cd}, p) = \left[ \exp\left( \frac{(p^2 + m_\phi^2)^{1/2}}{\zeta_\text{cd} T_\text{cd}} \right)  -1 \right]^{-1} \eqsp.
\label{eq:em_initial}
\end{align}
Here we assume that $\phi$ does not develop a chemical potential $\mu$ during freeze-out from the hidden sector thermal bath. A negligible chemical potential is naturally realised as long as $\phi$ decouples
from the plasma while relativistic, which applies to most of the parameter space we study. For the case of late decoupling however non-vanishing chemical potentials might well develop~\cite{Bringmann:2020mgx}. This will then typically translate to a {\it larger} number density $n_\phi$, implying that the bounds assuming $\mu=0$ are conservative in this case.

After decoupling, we assume that $\phi$ decays into two photons $\gamma$ or an electron-positron pair $e^\pm$.\footnote{In the following we will consider 1, 2 and 3 degrees of freedom in $\phi$, e.g.\ corresponding to real scalar, complex scalar and vector particles, respectively. As the Landau-Yang theorem forbids decays of on-shell vectors into two photons, one should think of the last case more broadly as 3 real scalars.} In our treatment we account for inverse decays as well as the spin-statistical factors for Bose enhancement and Pauli blocking. The corresponding Boltzmann equation can be written as
\begin{align}
\frac{\partial f_\phi(t, p)}{\partial t} - H(t)p\frac{\partial f_\phi(t, p)}{\partial p}= & \; \frac{1}{2E_\phi} \int \frac{\d^3 p_z}{(2 \pi)^3 2 E_z} \frac{\d^3 p_{\bar{z}}}{(2 \pi)^3 2 E_{\bar{z}}} \left| \mathcal{\overline{M}}_{\phi \rightarrow z \bar{z}} \right|^2 (2 \pi)^4 \delta^{(4)} (p - p_z - p_{\bar{z}}) \nonumber \\
&\times \left[ - f_\phi (1 \pm f_z) (1 \pm f_{\bar{z}}) + f_z f_{\bar{z}} (1+f_\phi) \right]\eqsp, \label{eq:Boltz_eq_gen}
\end{align}
where $H$ is the Hubble rate, $z \in \{ \gamma, e^-\}$, $\bar{z} \in \{ \gamma, e^+ \}$, $E_{z,\bar{z}} = \sqrt{m_{z, \bar{z}}^2 + p_{z, \bar{z}}^2}$, $\left| \mathcal{\overline{M}}_{\phi \rightarrow z \bar{z}} \right|$ is the matrix element averaged (summed) over the spins of all initial-state (final-state) particles (note that $\left| \mathcal{M}_{\phi \rightarrow z \bar{z}} \right| = \left| \mathcal{M}_{z \bar{z} \rightarrow \phi} \right|$), $f_{z,\bar{z}} = f_{z,\bar{z}} (t, p_{z,\bar{z}})$ are the corresponding distribution functions and the $+$ $(-)$ sign is used for photons (electrons/positrons). The matrix element can be related to the lifetime via~\cite{Zyla:2020zbs}
\begin{align}
\frac{1}{\tau_\phi} = \frac{\beta_z}{16 \pi m_\phi} \left| \mathcal{\overline{M}}_{\phi \rightarrow z \bar{z}} \right|^2\eqsp,
\end{align}
where $\beta_z = \sqrt{1-4 m_z^2 / m_\phi^2}$. As the SM decay products thermalise with a timescale much shorter than $\tau_\phi$ and $1/H$, their distribution functions are well described by Bose-Einstein and Fermi-Dirac distribution functions for bosons and fermions respectively. We use this to recast the Boltzmann equation in the form
\begin{align}
\frac{\partial f_\phi(t, p)}{\partial t} - H(t)p\frac{\partial f_\phi(t, p)}{\partial p}= -D_\y^\pm(t, p) \times \Big[ f_\phi(t, p) - \bar{f}_\phi(t, p) \Big]
\label{eq:em_solution_short}
\end{align}
with the equilibrium, i.e.\ Bose-Einstein, distribution function $\bar{f}_\phi$ with SM temperature $T$ and
\begin{align}
D_\y^\pm(t, p) = \frac{m_\phi}{E_\phi \tau_\phi} \left[ 1 + \frac{2T}{\beta_\y p} \ln\left( \frac{1 \mp \exp[-(E_\phi + \beta_\y p)/2T]}{1\mp\exp[-(E_\phi - \beta_\y p)/2T]} \right) \right]\bigg|_{T = T(t)}\eqsp,
\label{eq:em_D_term}
\end{align}
where the upper (lower) sign is used for decays into bosons (fermions),
and $E_\phi = \sqrt{p^2 + m_\phi^2}$ is the energy of $\phi$ for this momentum mode. Eq.~\eqref{eq:em_solution_short} can easily be generalised to decays into different SM final states by considering a linear superposition on the r.h.s.\ weighted with corresponding branching ratios.

The formal solution of eq.~\eqref{eq:em_solution_short} can be expressed as
\begin{align}
f_\phi(t, p) = &\; f_\phi\big(t_\text{cd}, pR(t)/R(t_\text{cd})\big) \exp\left( -\lint_{t_\text{cd}}^t \d t' \, D_\y^\pm\big(t', p R(t)/R(t')\big)\right) \nonumber \\[4mm]
& + \lint_{t_\text{cd}}^t \d t' \, \Theta\big(t', pR(t)/R(t')\big) \exp\left( -\lint_{t'}^t \d t'' \, D_\y^\pm\big(t'', p R(t)/R(t'')\big)\right)
\label{eq:em_solution}
\end{align}
with $\Theta(t, p) \ldefine D_\y^\pm \big(t,p\big) \times \bar{f}_\phi\big(t, p\big)$ and $R$ the scale factor. In practice $f_\phi(t, p)$ can be evaluated numerically using a Simpson rule in $\log (t/t_\mathrm{cd})$ for the first integral and a modified trapezoidal rule in log-log space for the second integral~\cite{Depta:2020wmr}. The phase-space distribution function is connected to the number and energy densities via the well-known relations
\begin{align}
n_\phi (t) &= g_\phi \int \frac{\d^3 p}{(2 \pi)^3} f_\phi (t, p)\eqsp, \\
\rho_\phi (t) &= g_\phi \int \frac{\d^3 p}{(2 \pi)^3} E_\phi f_\phi (t, p)\eqsp,
\end{align}
with $g_\phi$ the number of degrees of freedom of $\phi$.
To close the system of equations we need the Hubble rate as well as the ordinary differential equations for the time-temperature relation of the SM. The former is given by
\begin{align}
H(t) = \sqrt{\frac{8\pi G}{3}\left[ \rho_\sm(t) + \rho_\phi(t) \right]} \eqsp,
\label{eq:Hubble_em}
\end{align}
where $G$ is Newton's gravitational constant, $\rho_\sm$ is the SM energy density, and $\rho_\phi$ is the energy density of $\phi$. As the decay of $\phi$ leads to a heating of the SM thermal bath, the SM time-temperature relation becomes modified. Before and after neutrino decoupling, denoted by $\nu$d, it follows from the ordinary differential equations
\begin{align}
\frac{\d T}{\d t} &= -\frac{\dot{q}_\phi(t) + 3 \eq{H}(T)\big[\eq{\rho}_\elmnu(T) + \eq{\p}_\elmnu(T)\big]}{\d \eq{\rho}_\elmnu(T)/\d T} \qquad \text{for} \quad T \geq T_{\nu\d}\eqsp,
\label{eq:ode_temp_before_em} \\
\frac{\d T}{\d t} &= -\frac{\dot{q}_\phi(t) + 3 \eq{H}(T)\big[\eq{\rho}_\elm(T) + \eq{\p}_\elm(T)\big]}{\d \eq{\rho}_\elm(T)/\d T} \qquad \text{for} \quad T < T_{\nu\d} \eqsp,
\label{eq:ode_temp_after_em}
\end{align}
where $\rho$ denotes energy density, $P$ denotes pressure, and EM indicates that only electromagnetic degrees of freedom, i.e.\ photons, electrons, and positrons, are to be taken into account. The volume heating rate evaluates to
\begin{align}
\dot{q}_\phi(t) = -g_\phi \int \frac{\d^3 p}{(2\pi)^3} \, D_\phi(t, p) \Big[ f_\phi(t, p) - \bar{f}_\phi(t, p) \Big] \sqrt{p^2 + m_\phi^2}\eqsp,
\end{align}
which follows from integrating eq.~\eqref{eq:em_solution_short} over $\int \d^3 p \, E g_\phi /(2\pi)^3$.\footnote{However, we find that for numerical calculations it is best to determine $\dot{q}_\phi(t)$ directly from its definition $\dot{q}_\phi(t) = \dot{\rho}_\phi(t) + 3H(t)\big[ \rho_\phi(t) + \p_\phi(t)\big]$, since the numerical evaluation of $f_\phi(t, p) - \bar{f}_\phi(t, p)$ can easily lead to cancellation of significant digits.} Given that the reaction rate which keeps the neutrinos in thermal equilibrium scales as $T^5$, it is a good approximation to assume that neutrino decoupling happens instantaneously when $T^5 / H$ reaches the value it has at neutrino decoupling only considering the SM degrees of freedom, i.e.\ $T_{\nu \d}^5 / H (T_{\nu \d}) \simeq 1.62 \times 10^{-10} \, \mathrm{GeV}^{4}$ taking $T_{\nu \d}^\elmnu \simeq 1.4 \, \mathrm{MeV}$~\cite{Dolgov:2002wy, Bennett:2019ewm}.
After decoupling, the neutrino temperature $T_\nu$ redshifts as $1/R$. Note that around the exclusion line the injected entropy amounts to $\lesssim 1\%$ relative to the SM entropy, rendering possible corrections to the assumption of thermal equilibrium of neutrinos with the rest of the SM until their instantaneous decoupling negligible, even if there is heat transfer between $\phi$ and the SM around neutrino decoupling.

As detailed above we make the generic assumption that the decaying particle $\phi$ was in thermal equilibrium within the dark sector, which generically has a different temperature than the SM heat bath, corresponding to
$\zeta_\text{cd} \neq 1$. For sufficiently short lifetimes equilibration with the SM will naturally occur due to inverse decays. However, even for rather long lifetimes inverse decays can significantly impact
the abundance of $\phi$ if the initial temperature ratio is small, $\zeta_\text{cd} \ll 1$.
\begin{figure}
	\includegraphics[width=0.495\textwidth]{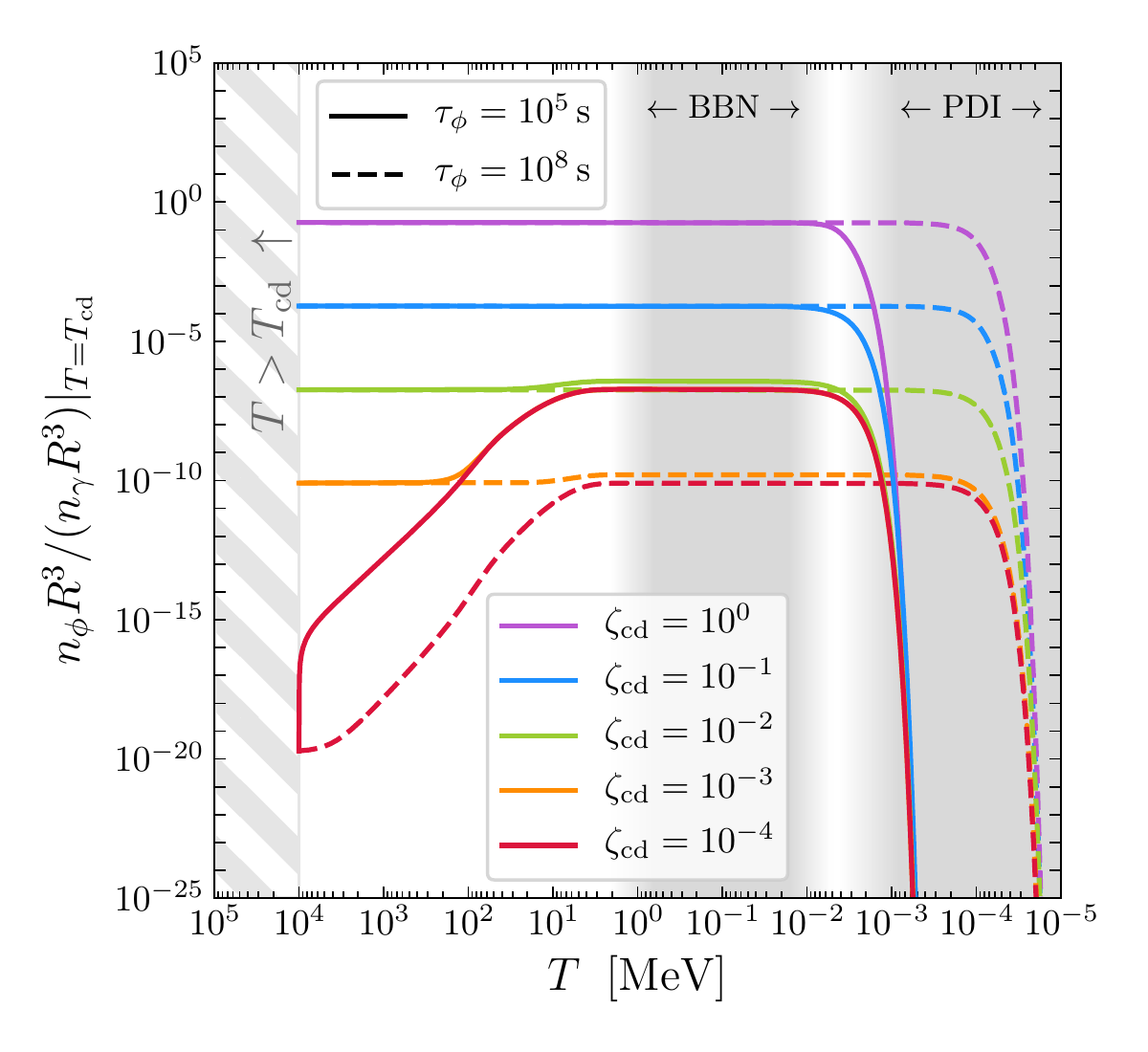}
	\includegraphics[width=0.495\textwidth]{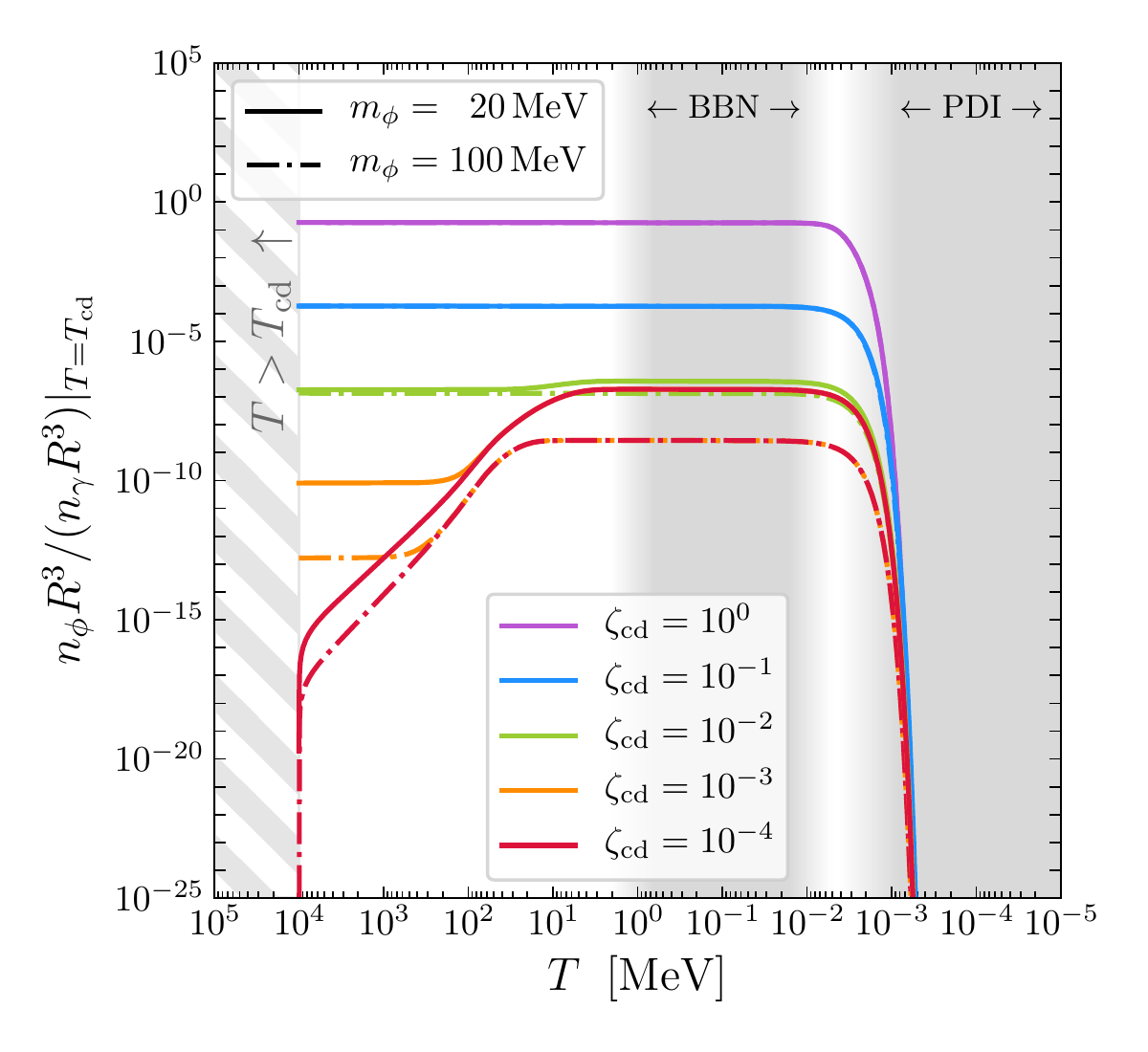}
	\caption{Left: Comoving number density of $\phi$ normalised to the photon number density at chemical decoupling as a function of the SM temperature $T$ for different values of $\zeta_\text{cd}$. We fix the chemical decoupling temperature to $T_\text{cd} = 10 \, \mathrm{GeV}$ and take  $m_\phi = 20 \, \mathrm{MeV}$, and $\tau_\phi = 10^5 \, \mathrm{s} ~ (10^8 \, \mathrm{s})$ in full (dashed).
	Right: Same, but for fixed lifetime, $\tau_\phi = 10^5 \, \mathrm{s}$ and $m_\phi = 20 \, \mathrm{MeV} ~ (100 \, \mathrm{MeV})$ in full (dash-dotted). We also indicate the temperature region relevant for BBN and photodisintegration (PDI).}
	\label{fig:nphi}
\end{figure}
In this case, there is a significant contribution to the abundance of $\phi$ due to `freeze-in' via inverse decays. To illustrate this, we show the comoving number density, normalised by the photon number density at chemical decoupling, for different values of $\zeta_\text{cd}$, $\tau_\phi$, and $m_\phi$ in figure~\ref{fig:nphi}, starting our calculation at chemical decoupling in the dark sector (assumed to happen when the SM temperature is $T_\text{cd} = 10 \, \mathrm{GeV}$) and assuming the phase-space distribution given in eq.~\eqref{eq:em_initial}.

In the left panel we show an example for the evolution of $n_\phi$ with $m_\phi = 20 \, \mathrm{MeV}$ and $\tau_\phi = 10^5 \, \mathrm{s}~ (10^8 \, \mathrm{s})$ in full (dashed). We observe that for large values of $\zeta_\text{cd}$ the comoving $n_\phi$ is practically constant before the decay sets in, with the initial abundance simply determined by the temperature ratio $\zeta_\text{cd}$. For small values of $\zeta_\text{cd}$ on the other hand we see that inverse decays significantly contribute to $n_\phi$, with a `freeze-in' contribution which scales like $\propto 1/\tau_\phi$, cf.\ eq.~\eqref{eq:em_D_term}.  For $\tau_\phi = 10^5 \, \mathrm{s}$ and $\zeta_\text{cd} = 10^{-4}$ this leads to a jump in the number density directly after chemical decoupling.\footnote{In principle, this implies that the number density of $\phi$ is already dominated by inverse decays at chemical decoupling and we therefore do not start our calculation with the appropriate phase-space distribution function. However, the largest effect of `freeze-in' via inverse decays occurs at temperatures $\sim \max (m_\phi, T(t=\tau_\phi))$, and this early contribution can hence be neglected.}
For $\tau_\phi = 10^8 \, \mathrm{s}$ the freeze-in contribution is correspondingly smaller and no jump is present. Inverse decays become ineffective at temperatures $\sim \max (m_\phi, T(t=\tau_\phi))$, thus leading to a constant comoving number density afterwards.
In the right panel we show the effect of different masses with $m_\phi = 20 \, \mathrm{MeV}$ (full) and $m_\phi = 100 \, \mathrm{MeV}$ (dash-dotted) for $\tau_\phi = 10^5 \, \mathrm{s}$.
We observe that the generated comoving number density becomes smaller with increasing $m_\phi$, as inverse decays become ineffective at temperatures below the mass.

\begin{figure}
    \hspace{-0.03\textwidth}
	\includegraphics[width=0.569\textwidth]{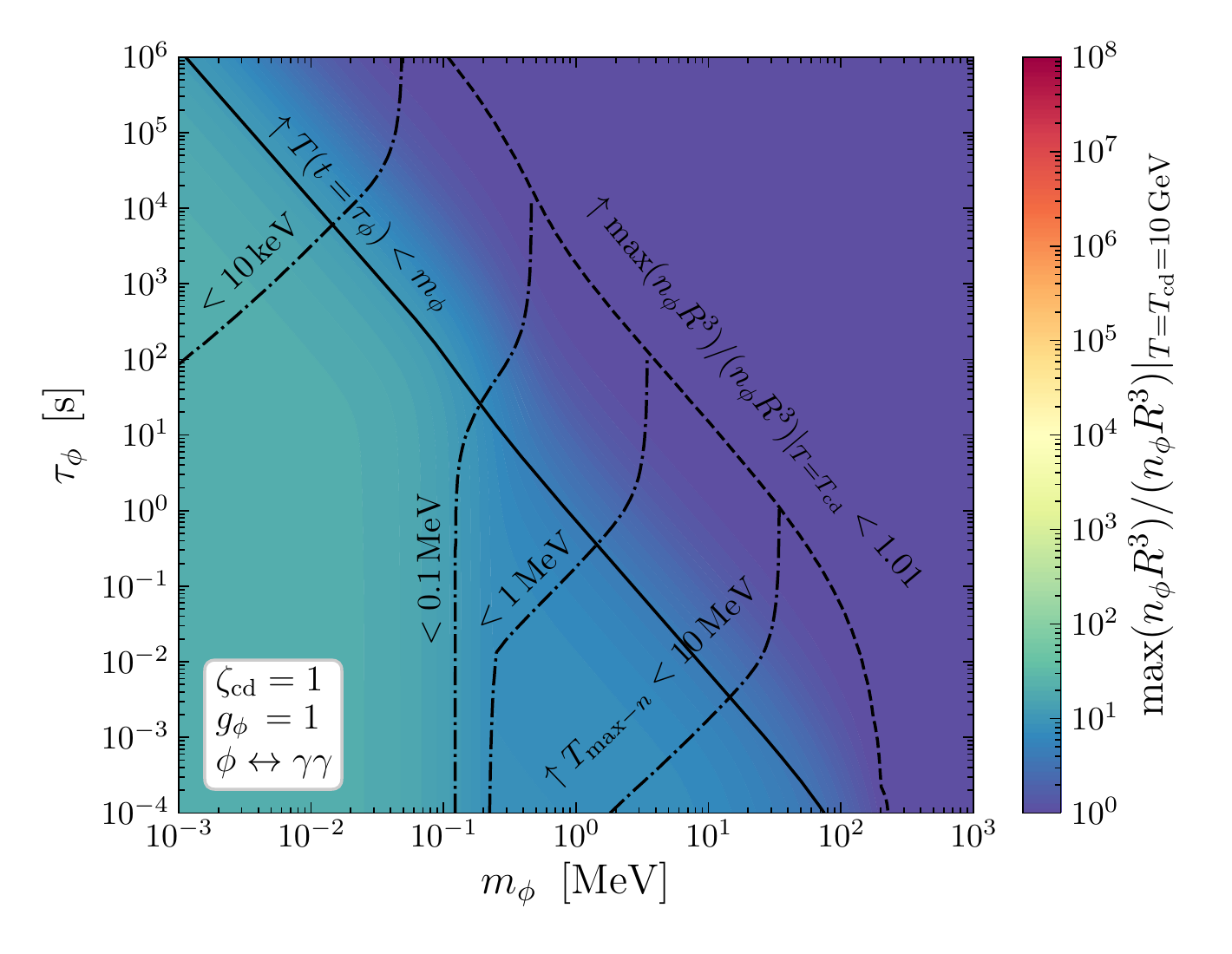}
	\hspace{-0.02\textwidth}
	\includegraphics[width=0.481\textwidth]{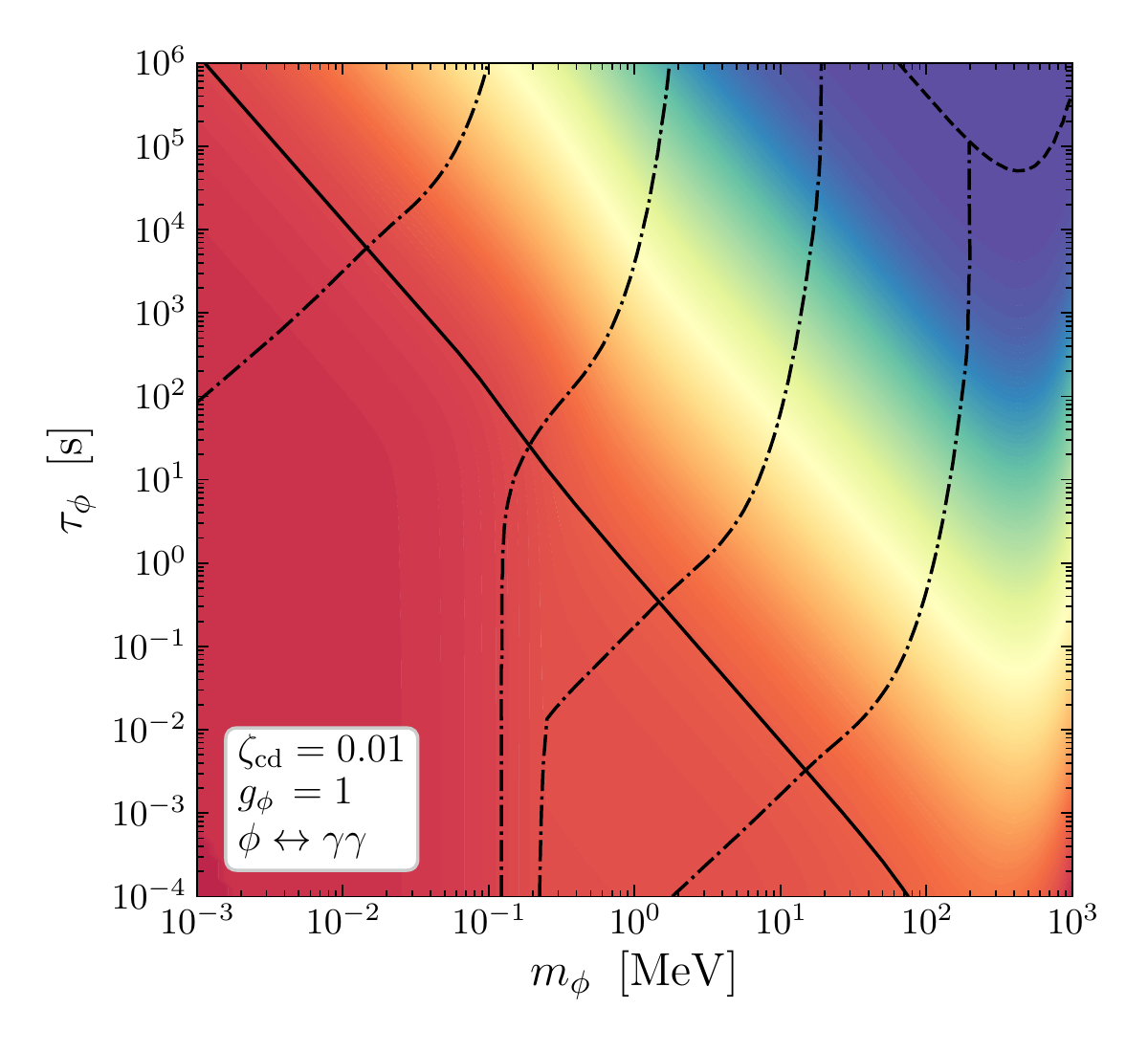}
	\caption{Maximal comoving $\phi$ number density normalised to its value at chemical decoupling assuming $g_\phi = 1$, decays into photons, and $\zeta_\text{cd} = 1~(0.01)$ left (right). We also indicate the contours, where the temperature at which the maximal comoving number density is first obtained is $T_{\mathrm{max}-n} < 10, \, 1, \, 0.1, \, 0.01 \, \mathrm{MeV}$ (dash-dotted lines), $T(t = \tau_\phi) = m_\phi$ (full line), and where the maximal comoving number density normalised to its value at chemical decoupling is $1.01$ (dashed line).}
	\label{fig:nphi_ratio}
\end{figure}
To further illustrate the importance of inverse decays we show the maximal comoving number density of $\phi$ normalised to its value at chemical decoupling in figure~\ref{fig:nphi_ratio} for two different values of $\zeta_\text{cd} = 1~(0.01)$ left (right). In general, the parameter space can be split into distinct regions, $T(t = \tau_\phi) \ll m_\phi$ and $T(t = \tau_\phi) \gg m_\phi$, and a transition region around $T(t = \tau_\phi) \sim m_\phi$. In the first region, $T(t = \tau_\phi) \ll m_\phi$, corresponding to the top right area, `freeze-in' via inverse decays is least effective as decays can already occur when inverse decays start to become relevant around $t \sim \tau_\phi$. Furthermore, inverse decays are kinematically suppressed by then. This implies that in the top right corner the maximal comoving number density normalised to its value at chemical decoupling is very close to one. Only once $T(t = \tau_\phi) / m_\phi$ becomes larger, inverse decays start to become more relevant as indicated by the dashed line below which they contribute more than $1 \%$ to the comoving number density.
Towards the bottom left corner $T(t = \tau_\phi) \gg m_\phi$ and $\phi$ thermalises with the SM as an additional relativistic degree of freedom, eventually becoming Boltzmann suppressed for $T (t \gg \tau_\phi) \ll m_\phi$. Hence, the maximal value of the comoving number density does not depend on $\tau_\phi$ for values of $\tau_\phi$ in this region. A dependence on $m_\phi$, however, remains as for smaller $m_\phi$ $\phi$ becomes non-relativistic and hence Boltzmann-suppressed at a later point. For small masses, a possible change in the effective number of relativistic SM degrees of freedom corresponding to an increase in temperature will further increase the comoving number density $n_\phi$ when $\phi$ has thermalised with the SM.
We also indicate the region where the maximal comoving number density is attained at a temperature of $T_{\mathrm{max}-n} < 10, \, 1, \, 0.1, \, 0.01 \, \mathrm{MeV}$ respectively by the dash-dotted lines.\footnote{We draw this line up to the dashed line as the comoving number density is approximately constant until decay above.}
Comparing the left and right panels, it is apparent that the effect of freeze-in via inverse decays is larger for $\zeta_\text{cd} = 0.01$, as the value of the number density at chemical decoupling is smaller by a factor of $10^6$. Hence, inverse decays can lead to a relative change in the comoving number density of $1 \%$ for larger values of $m_\phi$ and $\tau_\phi$ (dashed line) and the transition between the regions where inverse decays are not important ($T(t = \tau_\phi) \ll m_\phi$) to where $\phi$ thermalises ($T(t = \tau_\phi) \gg m_\phi$) starts for smaller values of $T(t = \tau_\phi) / m_\phi$. In the region, where $\phi$ fully thermalises with the SM, the final number density $n_\phi$ is, in fact, identical in both cases.
In the bottom left corner, we therefore observe a factor of $10^6$ between the two panels which simply reflects the different initial abundances at chemical decoupling.

Overall we see that the comoving number density of $\phi$ during the time of BBN and photodisintegration can be substantially different from the initial value at $T_\text{cd}$ due to (inverse) decays, in particular
for small values of $\zeta_\text{cd}$ where the initial abundance is suppressed. We also see that inverse decays can significantly change the number density $n_\phi$ even at rather small temperatures, as indicated by the dash-dotted lines. As a consequence the comoving number density of $\phi$ {\it is not comovingly constant} during BBN, implying that the evolution of $n_\phi$ has to be tracked carefully in order to correctly evaluate the resulting primordial element abundances.
In particular, it violates the commonly used assumption in many existing BBN studies of a comovingly constant $n_\phi$ before decay.
Let us stress that the effects of inverse decays only depend on the lifetime $\tau_\phi$ and are therefore completely model-independent. Regarding the effect of inverse decays during photodisintegration we note that for the relevant lifetimes, $\tau_\phi \gtrsim 10^4 \, \mathrm{s}$, the assumption of a comovingly constant $n_\phi$ is in fact justified unlike in the case for BBN. Nevertheless, it is of course crucial to include the effect of inverse decays when mapping these constraints to a number density $n_\phi$ at earlier times, e.g.\ at the time of freeze-out.
The cosmological evolution of $n_\phi$ discussed here should be kept in mind in later sections, when we show the resulting BBN limits for a given $\phi$ abundance at $T_\text{cd} = 10 \, \mathrm{GeV}$.

\section{Primordial element abundances}
\label{sec:BBN}

We use the latest recommendations for the observed abundances of $\mathcal{Y}_\mathrm{p}$ and $\text{D}/{}^1\text{H}$ from~\cite{Zyla:2020zbs} as well as ${}^3 \text{He}/\text{D}$ as an upper limit from~\cite{Geiss2003}:
\begin{align}
\mathcal{Y}_\mathrm{p} &= (2.45 \pm 0.03) \times 10^{-1} \label{eq:Yp_abundance} \eqsp,\\
\text{D}/{}^1\text{H} &= (2.547 \pm 0.025) \times 10^{-5} \label{eq:D_abundance} \eqsp,\\
{}^3\text{He}/\text{D} &= (8.3 \pm 1.5) \times 10^{-1} \label{eq:3HeH_abundance} \eqsp.
\end{align}
In addition the primordial lithium abundance is inferred to be \cite{Zyla:2020zbs}
\begin{align}
{}^7 \text{Li}/{}^1\text{H} &= (1.6 \pm 0.3) \times 10^{-10} \label{eq:Li_abundance} \eqsp.
\end{align}
Unlike the case for $\mathcal{Y}_\mathrm{p}$ and $\text{D}/{}^1\text{H}$ however, this measurement is roughly a factor three smaller than the value that is predicted by standard BBN
resulting in a significant tension. This is the well-known lithium problem.
In fact, some recent measurements in environments with metallicities below the Spite plateau indicate even smaller values of the lithium abundance~\cite{Aoki:2009ce}, possibly further strengthening the severity of this issue.
At present the origin of this discrepancy is still unresolved, but could be due to large astrophysical uncertainties, in particular due to stellar depletion~\cite{Korn:2006tv}.\footnote{A resolution of the discrepancy due to nuclear physics uncertainties is unlikely~\cite{Iliadis:2020jtc}.}
We therefore conservatively only take into account the limits from hydrogen, deuterium and helium-3 relative to deuterium in the next section. Nevertheless, another possibility which could resolve this tension could be the impact of new physics on the evolution of the different abundances which changes the theoretical prediction of lithium. We do discuss a scenario in which a decaying $\mathrm{MeV}$-scale particle significantly impacts the predicted lithium abundance in section~\ref{sec:li}.

To calculate the light element abundances we use a custom version of \textsc{AlterBBN~v1.4} \cite{Arbey:2011nf,Arbey:2018zfh} with replaced functions for $T(t)$, $T_\nu(t)$ and $H(t)$ using the results of the calculation described in section~\ref{sec:evo_phi}. The nuclear rate uncertainties are taken into account via the procedure detailed in~\cite{Hufnagel:2018bjp}.

For the baryon-to-photon ratio $\eta$ we employ the latest measurements from Planck~\cite{Aghanim:2018eyx}. Since the decay of a $\mathrm{MeV}$-scale particle can change the effective number of neutrinos $N_\mathrm{eff}$, we have to take into account the correlation between the best-fit values of $\eta$ and $N_\mathrm{eff}$ (see also~\cite{Millea:2015qra}). We follow the procedure detailed in~\cite{Depta:2020wmr}. Using the latest Planck data, i.e.\ the $95 \%$ confidence region ellipse (Planck TT,TE,EE+lowE+lensing+BAO) in the $\Omega_b h^2-N_\mathrm{eff}$ plane in figure~26 of~\cite{Aghanim:2018eyx}, gives
\begin{align}
\eta_{N_\mathrm{eff}} = \overline{\eta} + r \sigma_\eta \frac{N_\mathrm{eff} - \overline{N}_\mathrm{eff}}{\sigma_{N_\mathrm{eff}}}\eqsp,
\end{align}
where
\begin{align}
\overline{\eta} = 6.128 \times 10^{-10}, \quad \sigma_\eta = 4.9 \times 10^{-12}, \quad
\overline{N}_\mathrm{eff} = 2.991 , \quad \sigma_{N_\mathrm{eff}} = 0.169,\quad
r = \; 0.677,
\end{align}
to be used in the BBN calculation. As the dependence of $\text{D}/{}^1\text{H}$ on $\eta$ is relatively large, one has to propagate the uncertainty accordingly such that the total experimental uncertainty $\sigma_{\text{D}/{}^1\text{H}}^\text{exp}$ becomes
\begin{align}
\sigma_{\text{D}/{}^1\text{H}}^\text{eta} &= \left| \frac{\d (\text{D}/{}^1\text{H})}{\d \eta} \sigma_\eta \sqrt{1-r^2} \right|_{\eta = \eta_{N_\mathrm{eff}}} \approx 0.024 \times 10^{-5} \eqsp, \\
\sigma_{\text{D}/{}^1\text{H}}^\text{exp} &= \sqrt{ \left( \sigma_{\text{D}/{}^1\text{H}}^\text{obs} \right)^2 + \left( \sigma_{\text{D}/{}^1\text{H}}^\text{eta} \right)^2} \approx 0.035 \times 10^{-5}\eqsp.
\end{align}
For the other abundances this procedure is not necessary as the observational uncertainty is much larger than the one due to the uncertainty of $\eta$.

\subsection{Photodisintegration}

The effect of photodisintegration on the primordial element abundances are calculated with the code \texttt{ACROPOLIS}~\cite{Depta:2020mhj},  which
is released in parallel to this article. For details of our treatment we therefore refer to~\cite{Hufnagel:2018bjp,Depta:2020mhj}.
Note that in the part of parameter space where photodisintegration has a considerable effect, i.e.\ for $m_\phi \gtrsim 3 \, \mathrm{MeV}$, $\tau_\phi \gtrsim 10^4 \, \mathrm{s}$, the effect of spin-statistical factors and inverse decays can be neglected for calculating the non-thermal part of the photon spectrum as the particle $\phi$ is highly non-relativistic when it decays.
Nevertheless, inverse decays can be very relevant in setting the abundance of $\phi$, in particular for small temperature ratios $\zeta_\text{cd}$.

\section{Results}
\label{sec:results}
\subsection{Constraints for freeze-out with $\mathbf{\zeta_\text{cd} =\;} 1$}
\label{sec:const_fo}

\begin{figure}[H]
\vspace{-4mm}
	\includegraphics[width=0.495\textwidth]{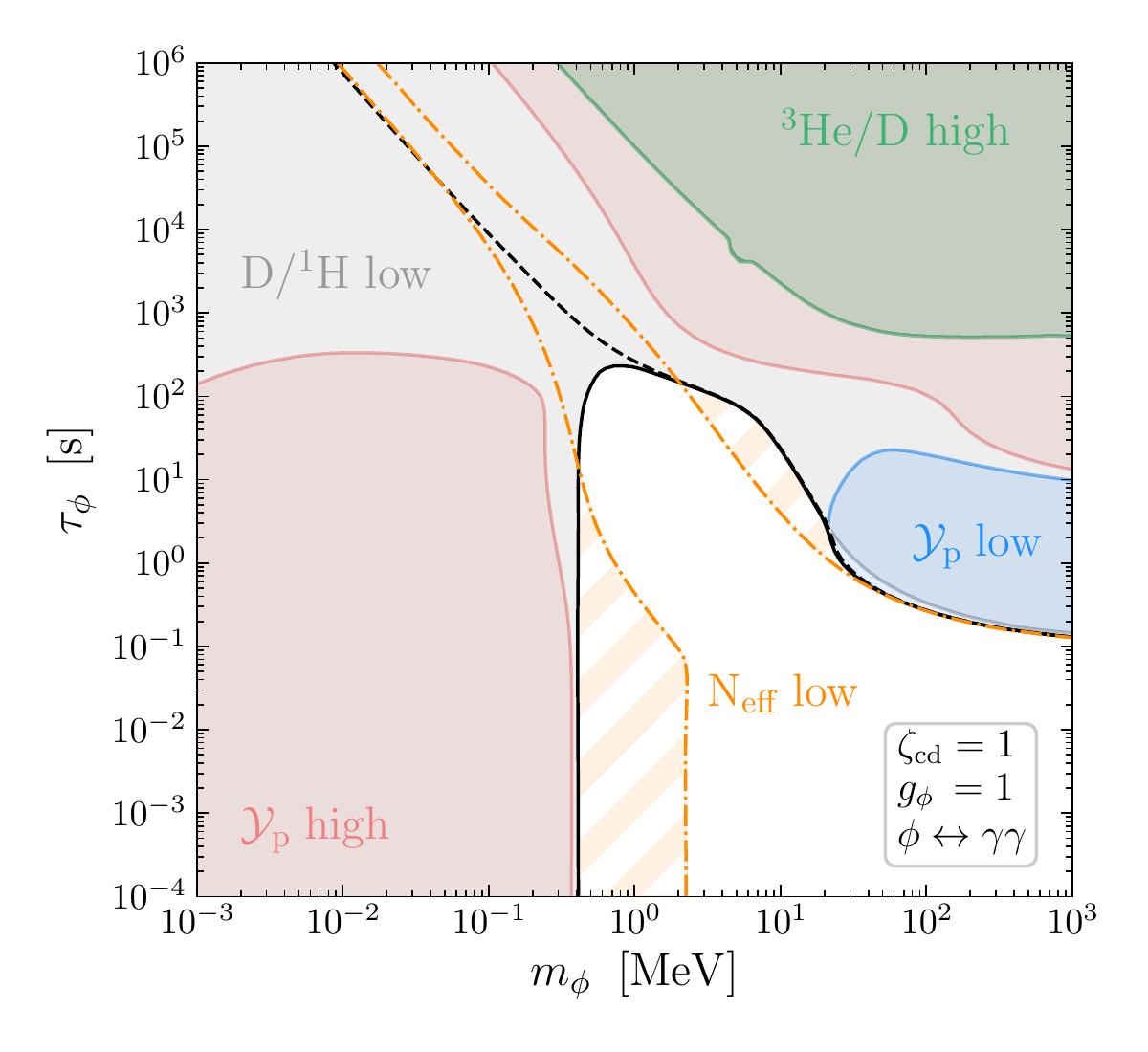}
	\includegraphics[width=0.495\textwidth]{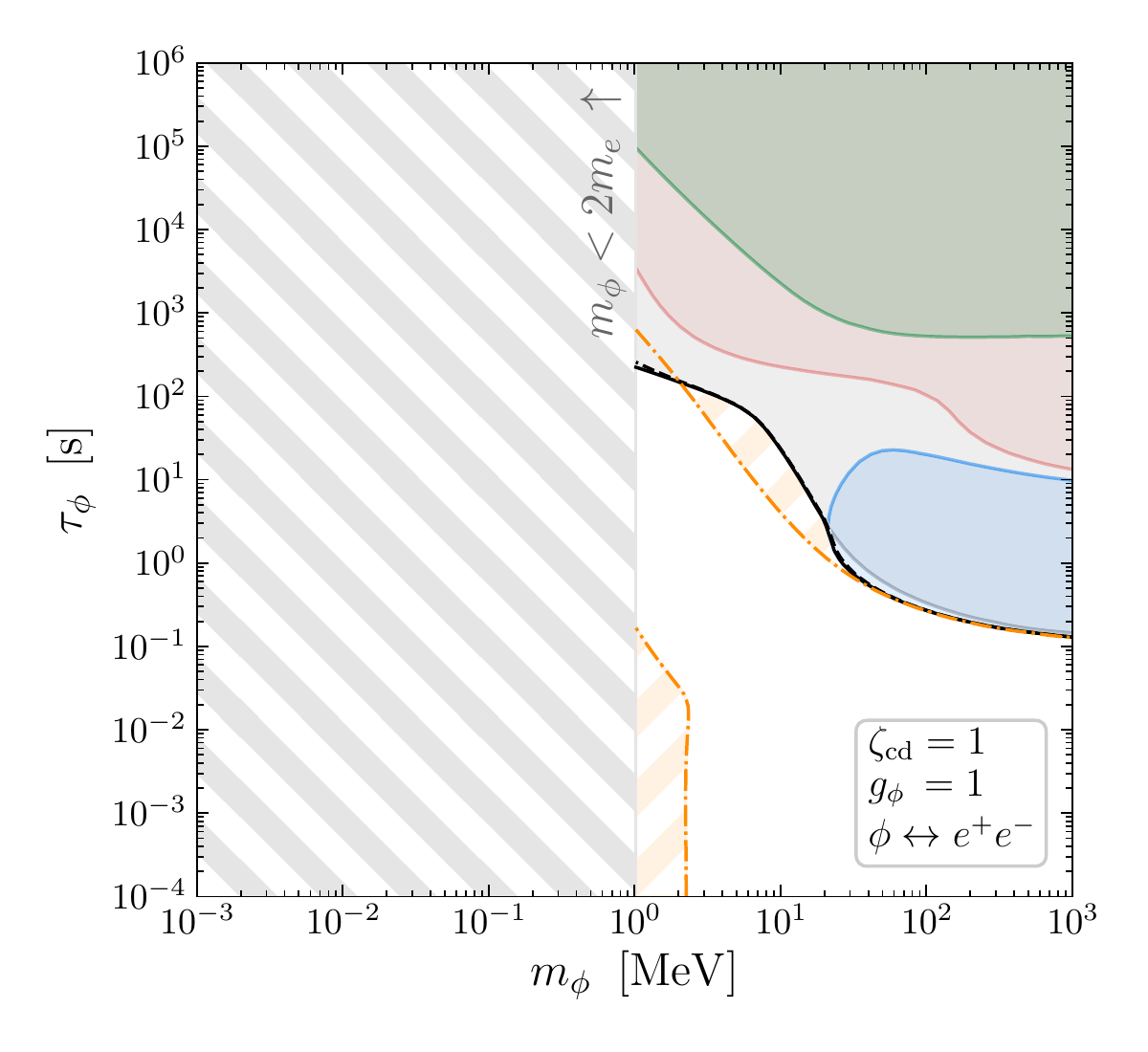}
	\includegraphics[width=0.495\textwidth]{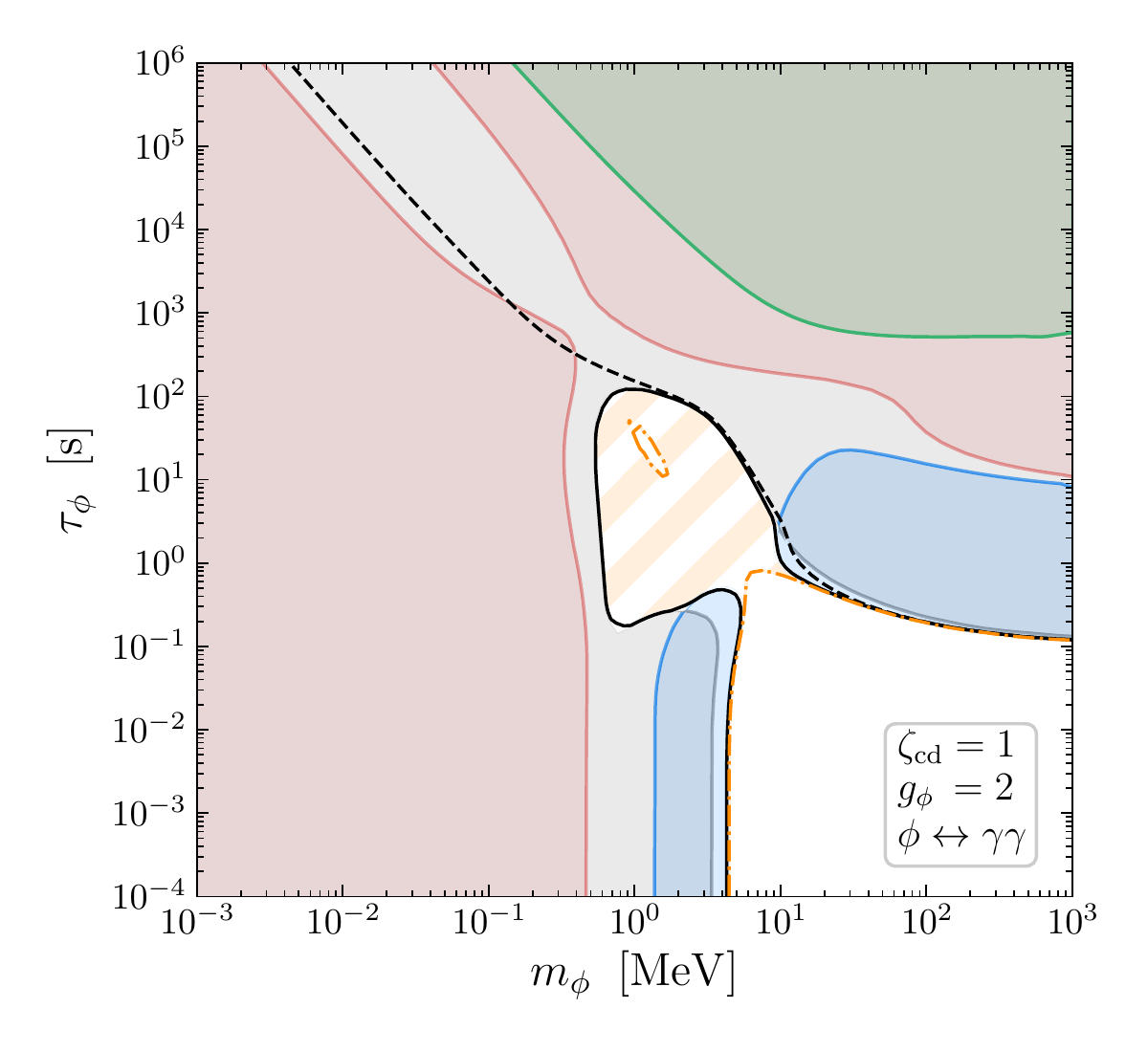}
	\includegraphics[width=0.495\textwidth]{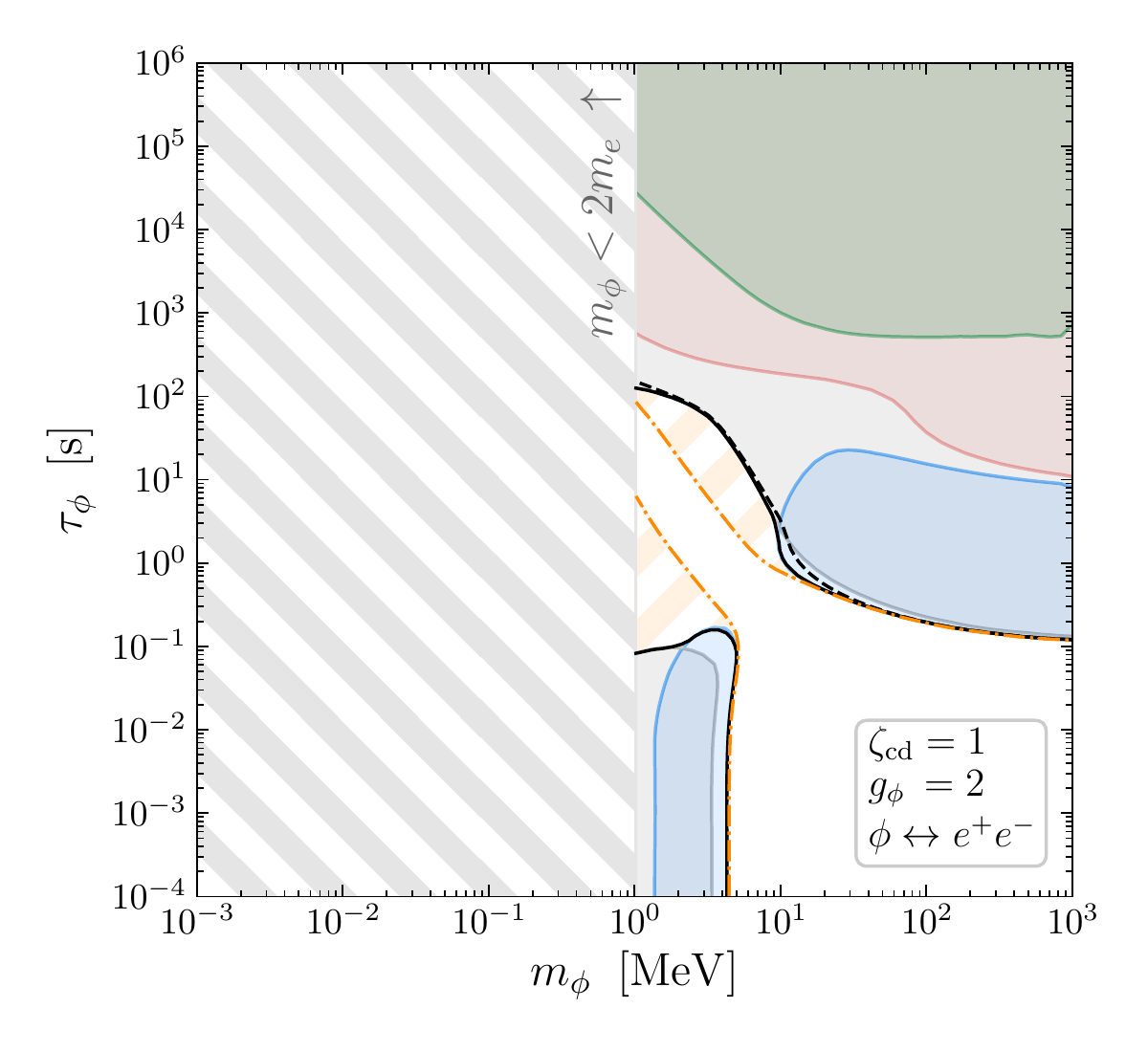}
	\includegraphics[width=0.495\textwidth]{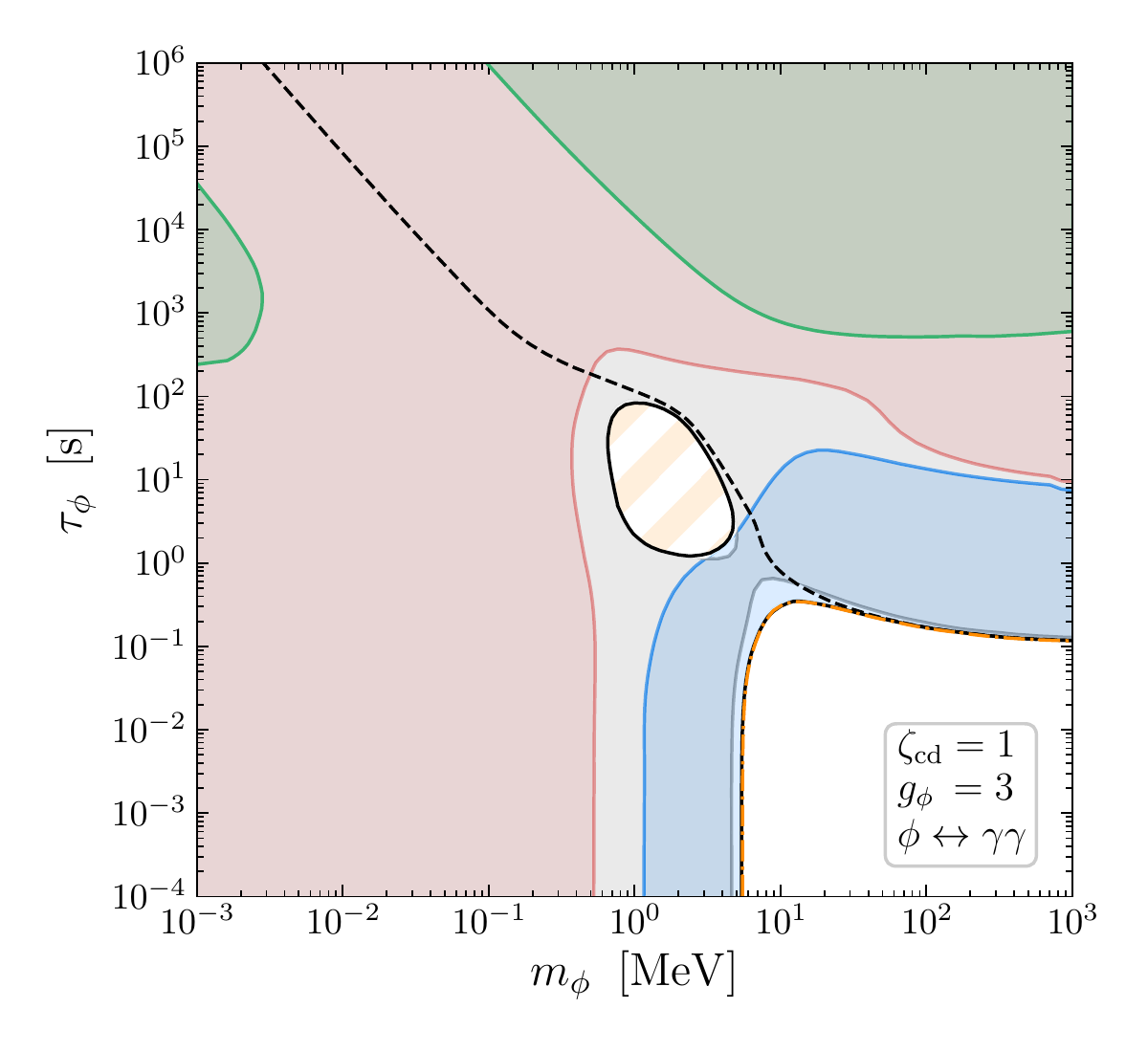}
	\includegraphics[width=0.495\textwidth]{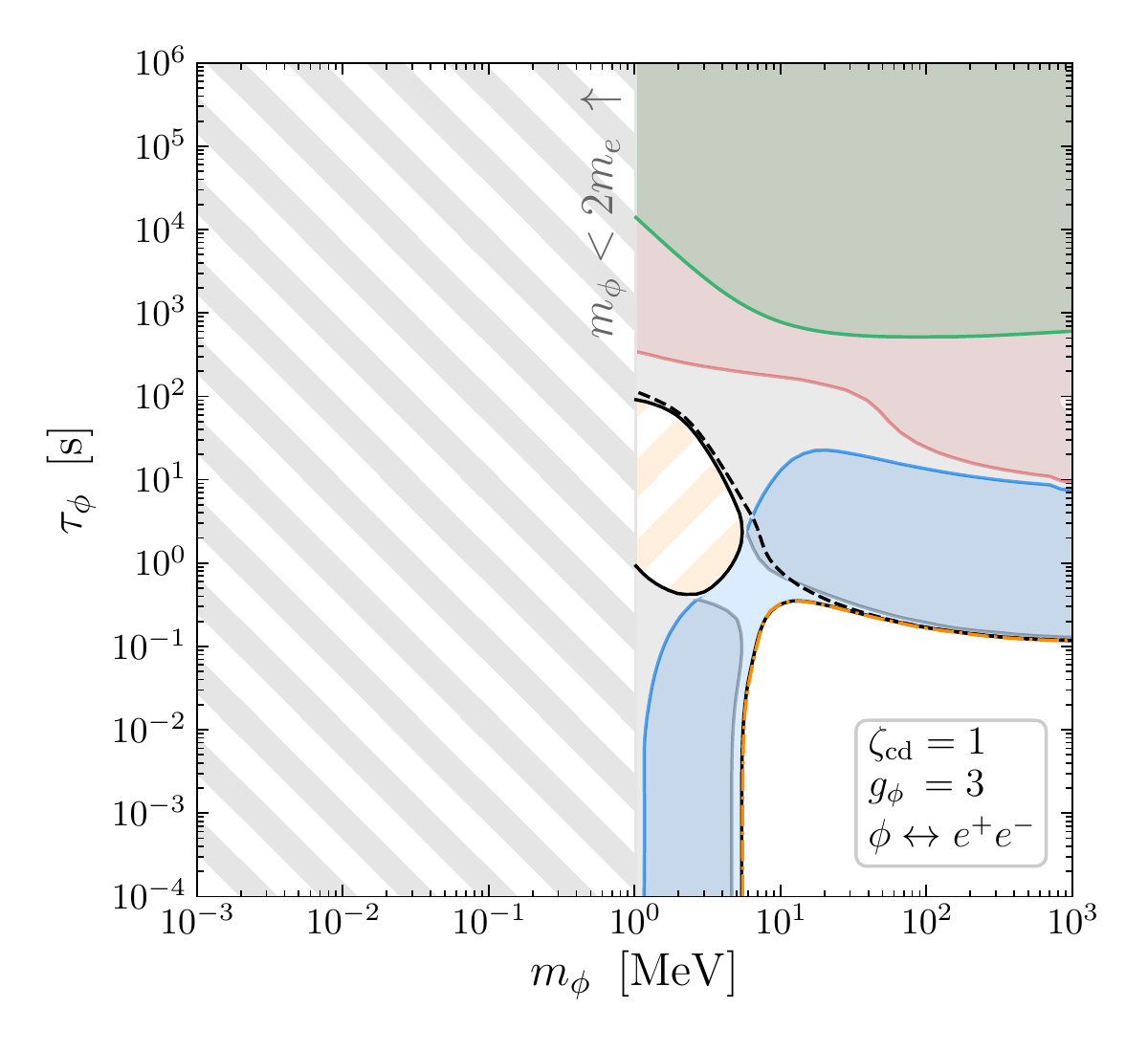}
\vspace{-7mm}
	\caption{95\% C.L.\ constraints in $m_\phi - \tau_\phi$ plane for $g_\phi=1$, 2 and 3 degrees of freedom respectively, decaying into two photons (left) or electron-positron pairs (right) assuming $T_\text{cd} = 10 \, \mathrm{GeV}$ and $\zeta_\text{cd} = 1$. The limits from individual observables are shown separately: primordial deuterium abundance (grey), helium-4 mass fraction $\mathcal{Y}_\text{p}$ (red high, blue low), helium-3 abundance normalised by deuterium (green), and Planck $N_\mathrm{eff}$  (orange, dash-dotted). The overall 95\% C.L.\ BBN limit is given by the black full line. For comparison we also show the overall limit neglecting inverse decays and spin statistics (black dashed).}
\label{fig:xi_1}
\end{figure}

Let us start with the case where the temperatures in the hidden sector and the SM at chemical decoupling are the same, $\zeta_\text{cd}=1$.
In figure~\ref{fig:xi_1} we show from top to bottom constraints in the $m_\phi - \tau_\phi$ plane for 1, 2 and 3 degrees of freedom respectively, decaying into two photons (left) or electron-positron pairs (right) with $T_\text{cd} = 10 \, \mathrm{GeV}$. The overall 95\% C.L.\ BBN limit is given by the black full line\footnote{For simplicity we in fact use the envelope of the individual 95\% C.L.\ constraints for the overall 95\% C.L.\ BBN limit, thus not taking into account any correlations, which leads to a somewhat more conservative limit.} while the dash-dotted orange line gives the limit from the CMB. For comparison, we also show the would-be overall limit when inverse decays and spin-statistics are neglected, which is indicated by the black dashed line. It can clearly be seen that including inverse decays leads to significantly stronger constraints in particular for
small masses and small lifetimes. To obtain an intuitive understanding of this result, we remind the reader that the full parameter space can essentially be divided into two phenomenologically distinct regions,  $T(t=\tau_\phi) \gg m_\phi$ (bottom left region) and{{\parfillskip0pt\par}}
\noindent $T(t=\tau_\phi) \ll m_\phi$ (top right region), cf.\ the discussion around figure~\ref{fig:nphi_ratio} above.
In particular, as can also be seen in figure~\ref{fig:nphi_ratio}, inverse decays do not appreciably affect the abundance of $\phi$ for $T(t=\tau_\phi) \ll m_\phi$ while they significantly add to the abundance for
$T(t=\tau_\phi) \gg m_\phi$.

For $T(t = \tau_\phi) \ll m_\phi$, i.e.\ $m_\phi \gtrsim 10 \, \mathrm{MeV}$, $\tau_\phi \gtrsim 0.1 \, \mathrm{s}$, the limits therefore resemble the results from \cite{Hufnagel:2018bjp}.
Here $\phi$ typically becomes non-relativistic between decoupling and its decay, leading to strong constraints due to the enhanced energy density if the decays happen
after the on-set of BBN (i.e.\ for $\tau_\phi \gtrsim 0.1\,\mathrm{s}$).

For $T(t=\tau_\phi) \gg m_\phi$, i.e.\ for $m_\phi \lesssim 10 \, \mathrm{MeV}$, $\tau_\phi \lesssim 0.1 \, \mathrm{s}$, $\phi$ thermalises with the SM at a time where the SM temperature is still large compared to the mass of $\phi$, i.e.\ $m_\phi \ll T$.
$\phi$ therefore behaves as one or more additional relativistic SM degrees-of-freedom and acquires a large thermal abundance, which is strongly constrained.
To evaluate this effect consistently, it is crucial to take into account the spin-statistical factors, implying a Bose-Einstein phase-space distribution of $\phi$.
To avoid tension with observations, $\phi$ must become non-relativistic sufficiently early so that the number density $n_\phi$ is already Boltzmann suppressed.
This implies that in this region there is a lower bound on the mass $m_\phi$ which depends mainly on the number of degrees-of-freedom $g_\phi$ resembling the limits from~\cite{Boehm:2013jpa, Nollett:2013pwa, Depta:2019lbe,Sabti:2019mhn}.

Regarding the limit from the CMB, it is important to note that for large regions of parameter space we consider, $\phi$ vanishes from the thermal bath after neutrino decoupling. It therefore heats up the photon bath and correspondingly leads to a {\it decreased} value of $N_\mathrm{eff}$ (as the neutrinos are already decoupled). The corresponding CMB limit (cf.\ orange dashed line), however, could straight-forwardly be circumvented by including additional dark radiation.

Comparing decays into photons (left panels) and electron-positron pairs (right panels) the major difference is that for $e^+/e^-$ the decay is kinematically forbidden for $m_\phi < 2 m_e$. In the rest of parameter space the limits are qualitatively similar, but small quantitative differences arise due to the different spin-dependent collision terms, cf.\ eq.~\eqref{eq:em_D_term}, especially when $m_\phi \sim 2 m_e$.
Comparing the limits for different numbers of degrees-of-freedom, we note that the constraints become increasingly more stringent due to the correspondingly enhanced energy densities.

In figure~\ref{fig:xi_1} we assume that $\phi$ chemically decoupled from the DS at $T_\text{cd} = 10 \, \mathrm{GeV}$.
Let us now briefly discuss how the limits will change if $\phi$ freezes out at another temperature $T_\text{cd}$.
In the left panel of figure~\ref{fig:Tcd} we show the effect of varying $T_\text{cd}$ for $\zeta_\text{cd}=1$.
We observe that as long as $T_\text{cd}$ is {\it above} the temperature of the QCD phase transition, $T_\text{QCD} \simeq 200 \, \mathrm{MeV}$, the limits are rather insensitive to changes in $T_\text{cd}$.
This is to be expected as the difference in the number densities $n_\phi$ when varying $T_\text{cd}$ only depends on the relativistic degrees of freedom $g_{*s}$ as long as the freeze-out
happens while $\phi$ is still relativistic. For $m_\phi \lesssim 0.4 \, \mathrm{MeV}$ inverse decays in fact lead to a thermalisation of the two sectors and the constraints are completely independent of $T_\text{cd}$.
For values of $T_\text{cd}$ {\it below} the QCD phase transition two effects come into play: For sufficiently small masses $m_\phi$ the freeze-out still happens while
$\phi$ is relativistic and -- assuming that the dark and visible temperatures are the same, $\zeta_\text{cd}=1$ -- the number density $n_\phi$ is relatively larger compared to the case where
only the SM temperature is increased due to the QCD phase transition. Correspondingly we see that constraints become stronger for small $m_\phi$.
For larger values of $m_\phi$, on the other hand, the number density $n_\phi$ is already Boltzmann suppressed at $T_\text{cd}$ and the constraints become significantly weaker.

Let us finally comment that there are some scenarios which cannot directly be mapped onto our framework. Nevertheless, our results still give a very good estimate of the constraints which can be expected.
For example, our limits also roughly apply to  ALPs with $\mathrm{MeV}$-scale masses and similar lifetimes as considered in~\cite{Millea:2015qra,Depta:2020wmr}.
For an ALP coupled to two photons, Primakoff interactions will establish thermal equilibrium with the SM for sufficiently large reheating temperatures, so $\zeta_\text{cd}=1$.
The main difference is that in the more model-independent framework considered here we take the SM temperature at chemical decoupling as a {\it free} parameter, whereas in the case of ALPs the decoupling temperature $T_\mathrm{cd}$ is not constant but depends on the other parameters. Nevertheless our result for $\zeta_\text{cd}=1$ and $T_\text{cd} = 10 \, \mathrm{GeV}$ are rather close to the actual result for
ALPs, which can be seen by comparing to figure 3 in~\cite{Depta:2020wmr}:
In the region, where $\phi$ thermalises the limits do not depend on $T_\text{cd}$ and are therefore identical, while in the other parts of parameter space the differences are small.

\subsection{Constraints for freeze-out with $\zeta_\text{cd} \neq 1$}

\begin{figure}
	\includegraphics[width=0.495\textwidth]{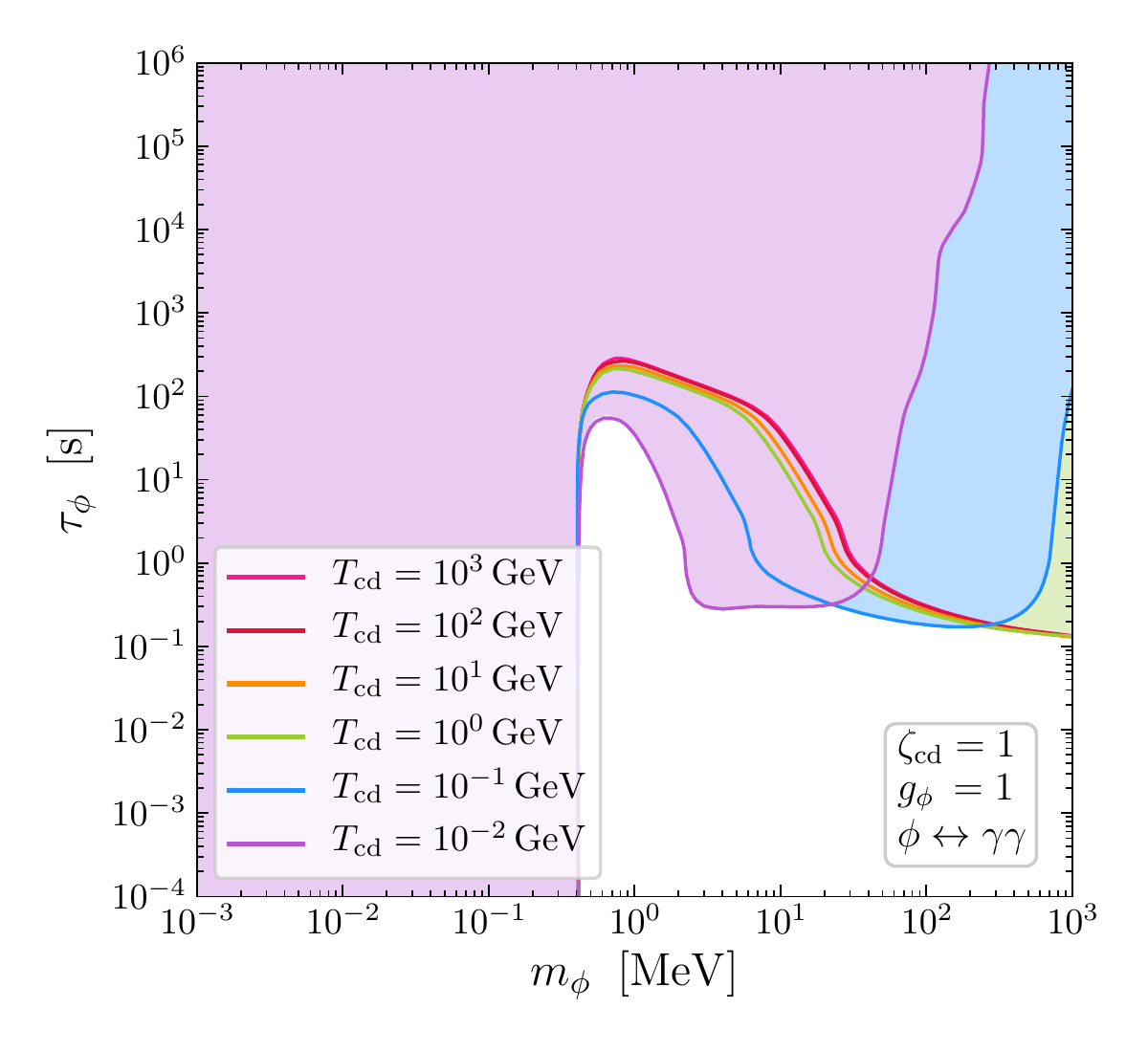}
	\includegraphics[width=0.495\textwidth]{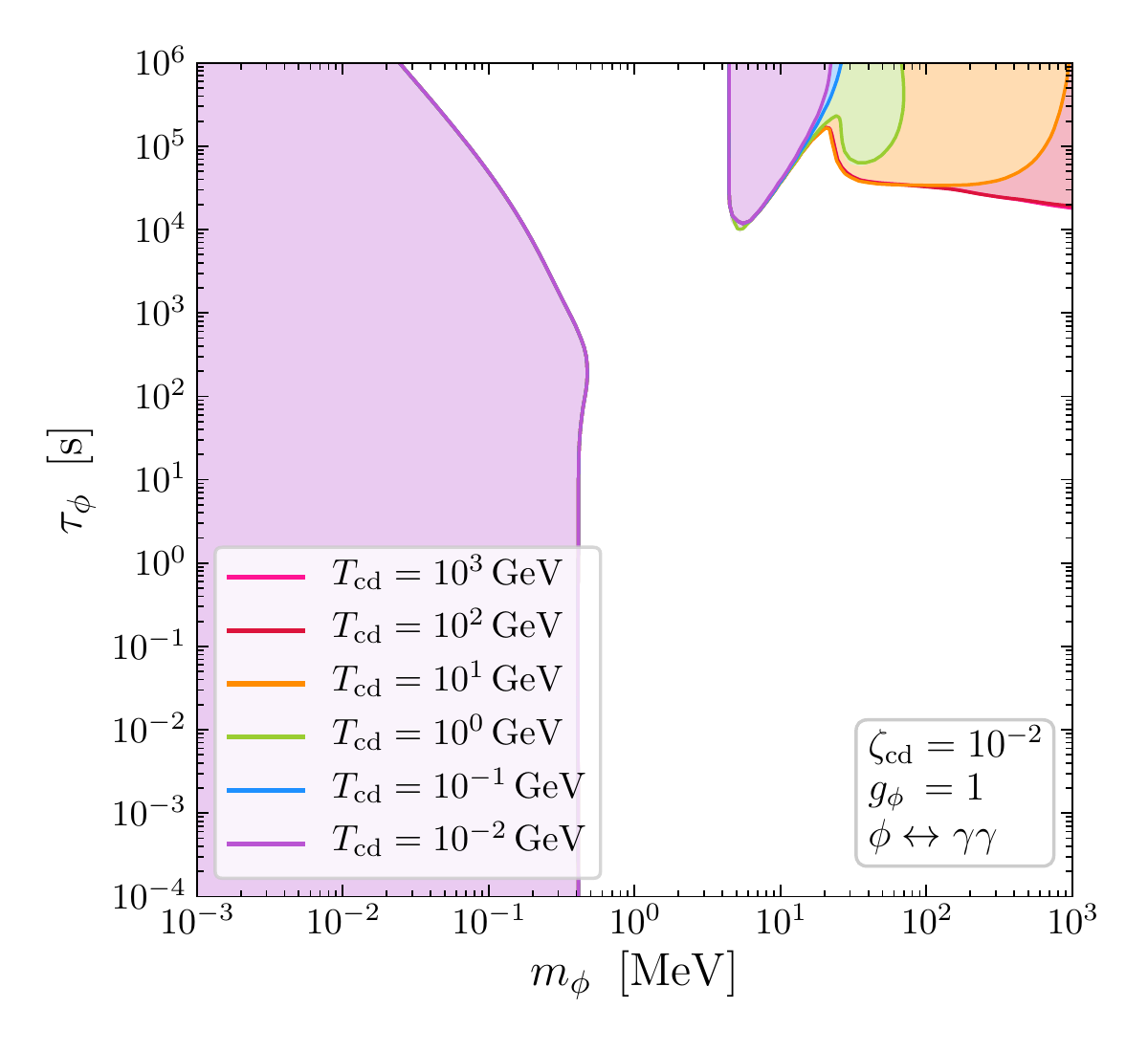}
	\caption{Overall 95\% C.L.\ BBN constraint (full) in the $m_\phi - \tau_\phi$ plane for $g_\phi=1$ and decays into two photons for different values of $T_\text{cd}$,
	fixing $\zeta_\text{cd} = 1$ (left) as well as $\zeta_\text{cd} = 0.01$  (right).}
	\label{fig:Tcd}
\end{figure}

In this section we discuss how the limits change when we allow for a temperature ratio at freeze-out different from unity, $\zeta_\text{cd} \neq 1$.
Let us start with a brief discussion of how a variation of the freeze-out temperature $T_\text{cd}$ changes the limits in this case.
In the right panel of figure~\ref{fig:Tcd} we show the effect of varying $T_\text{cd}$ for a rather small DS temperature, $\zeta_\text{cd}=0.01$.
We see that for $m_\phi \lesssim 0.4 \, \mathrm{MeV}$ the constraints are again completely independent of $T_\text{cd}$ due to the thermalisation of the two sectors.
For larger masses the abundances are so small that they can only be constrained via photodisintegration, implying only the region with{{\parfillskip0pt\par}}
\begin{figure}[H]
	\includegraphics[width=0.495\textwidth]{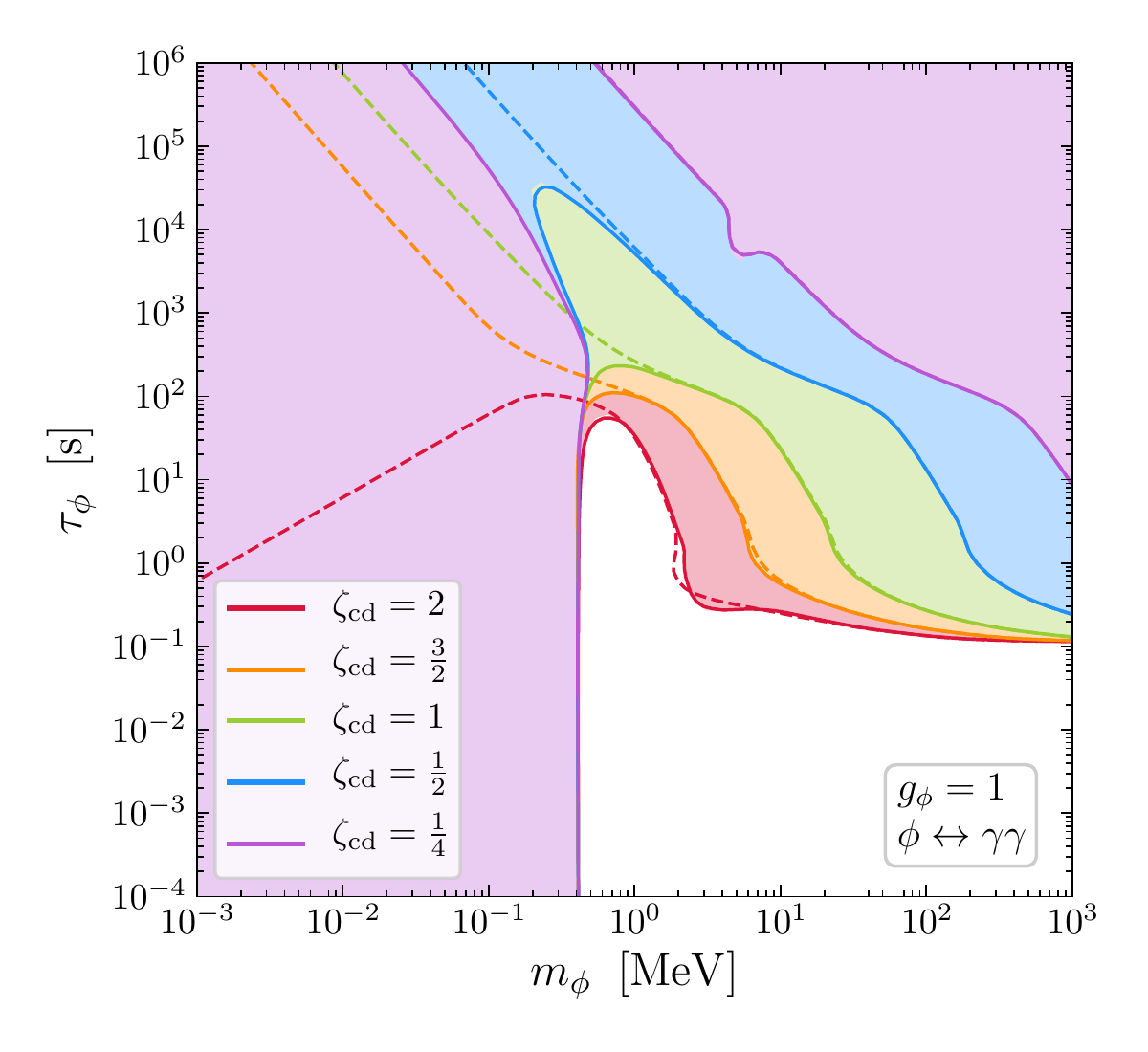}
	\includegraphics[width=0.495\textwidth]{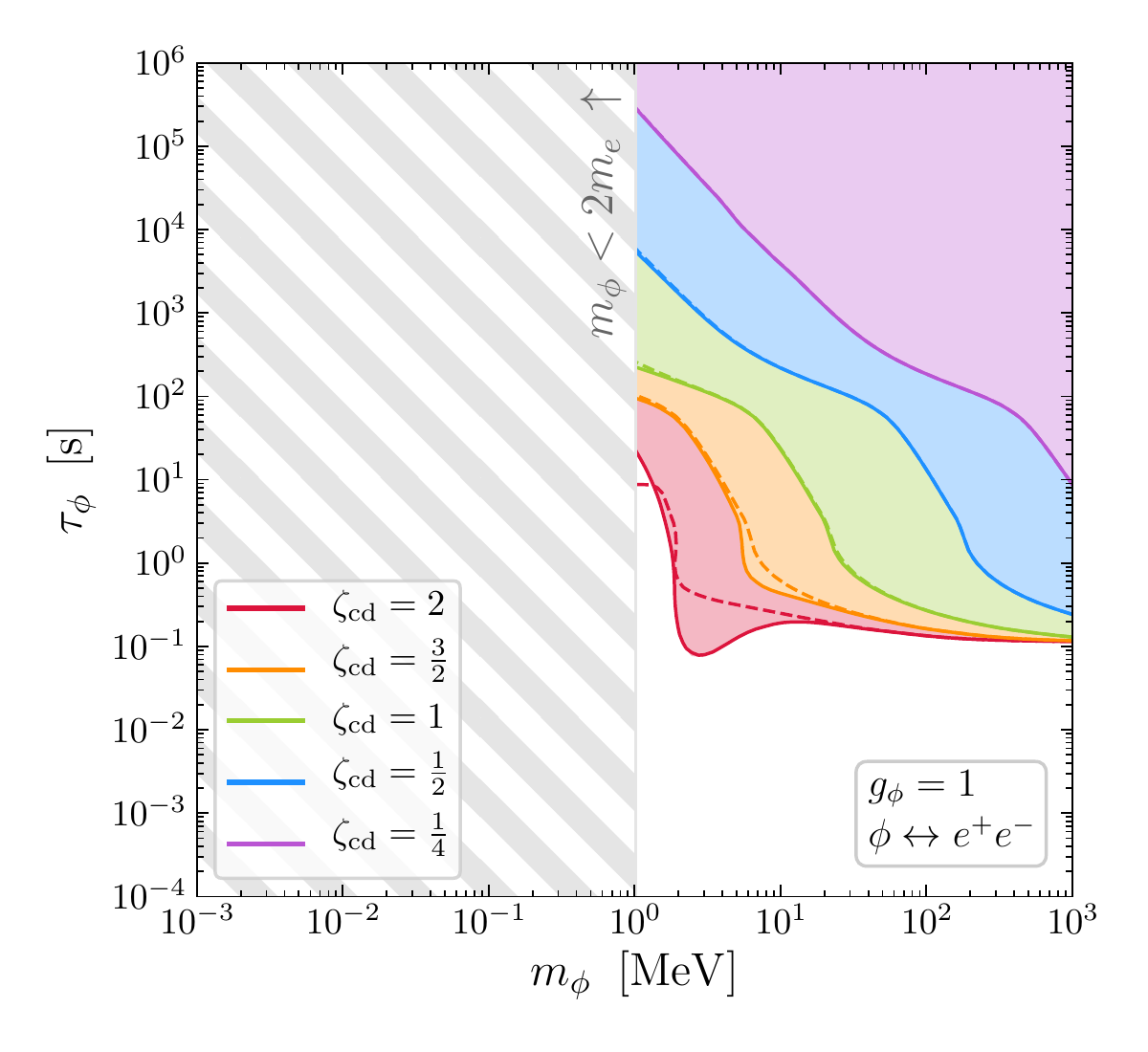}
	\includegraphics[width=0.495\textwidth]{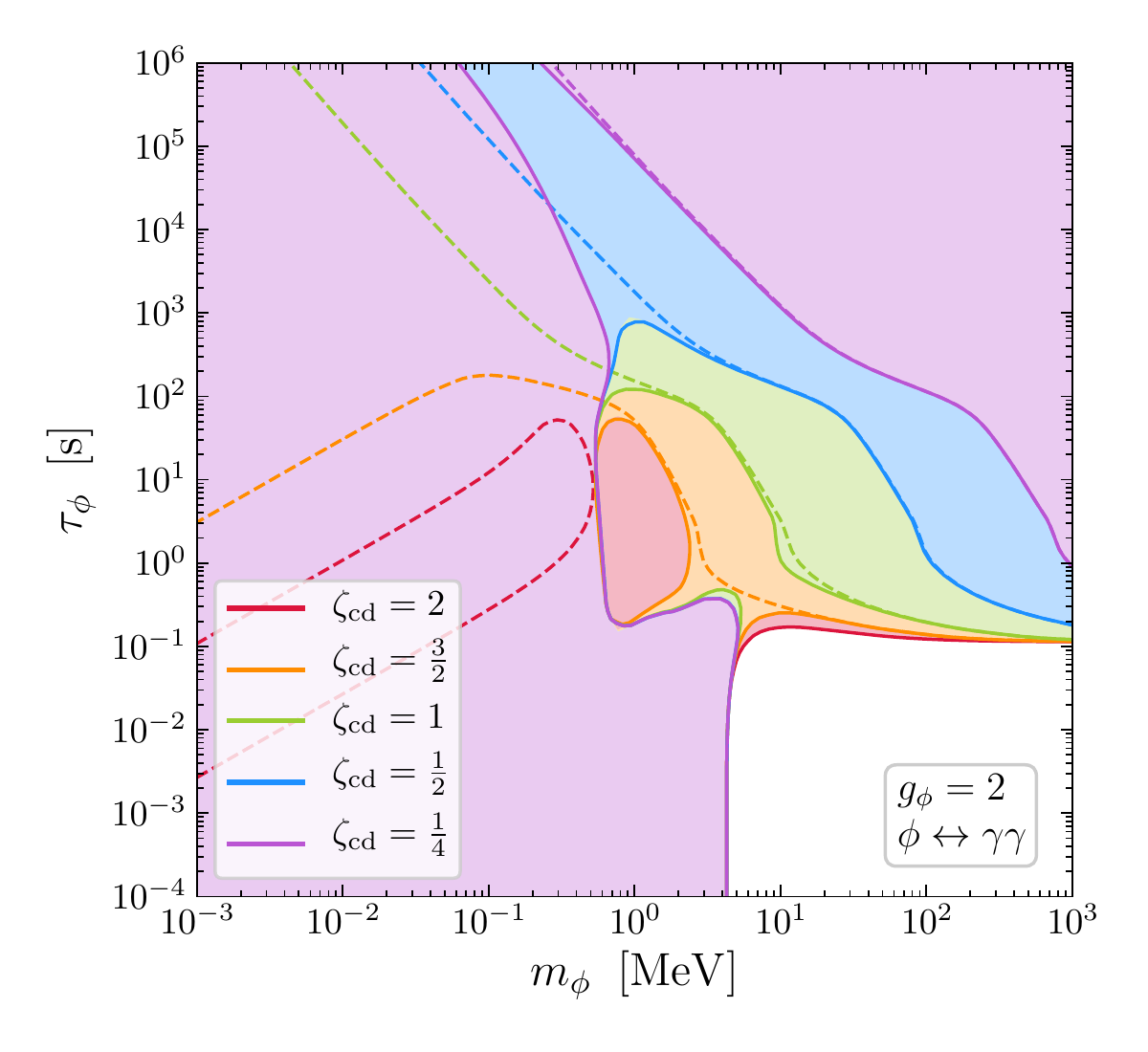}
	\includegraphics[width=0.495\textwidth]{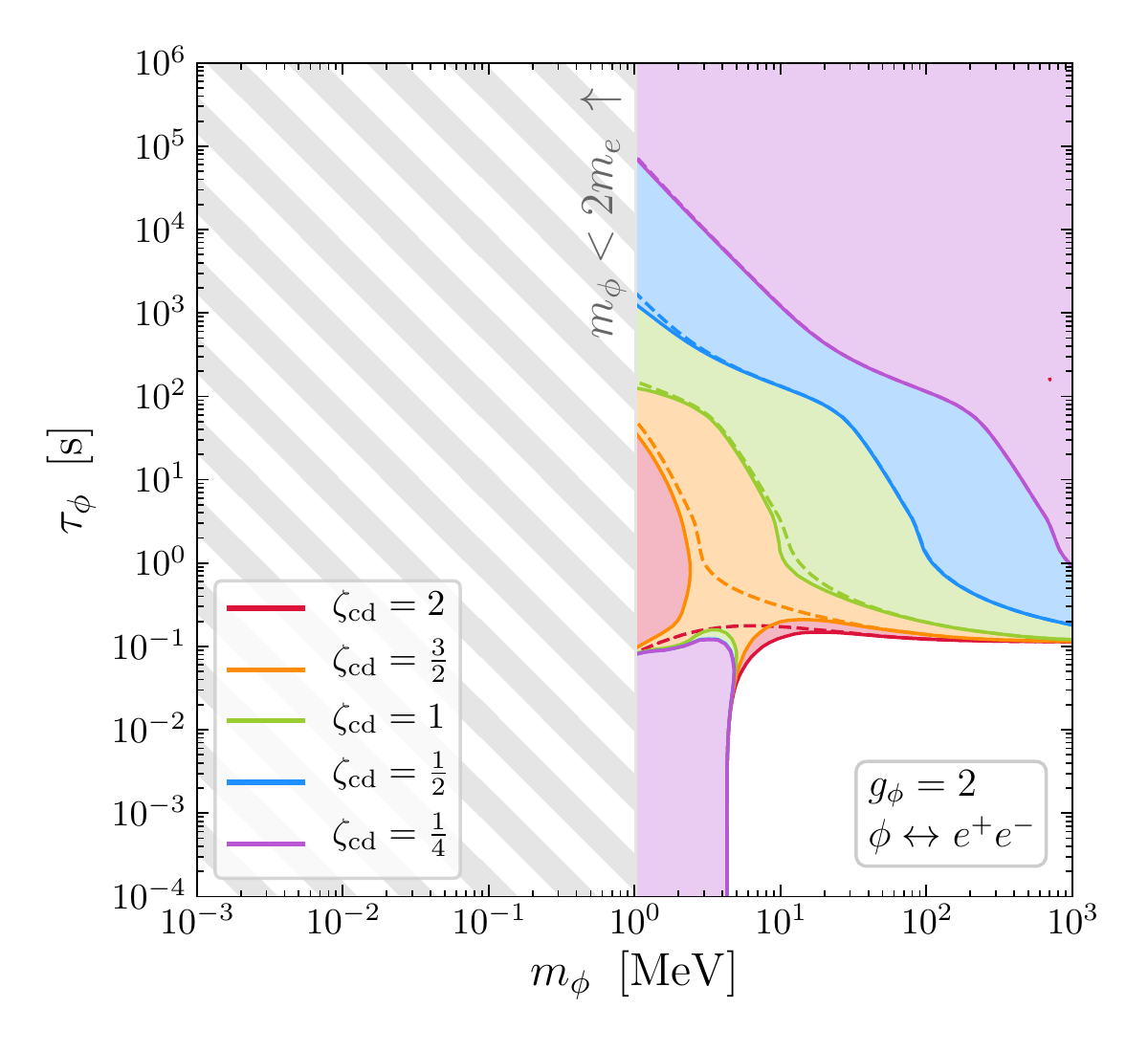}
	\includegraphics[width=0.495\textwidth]{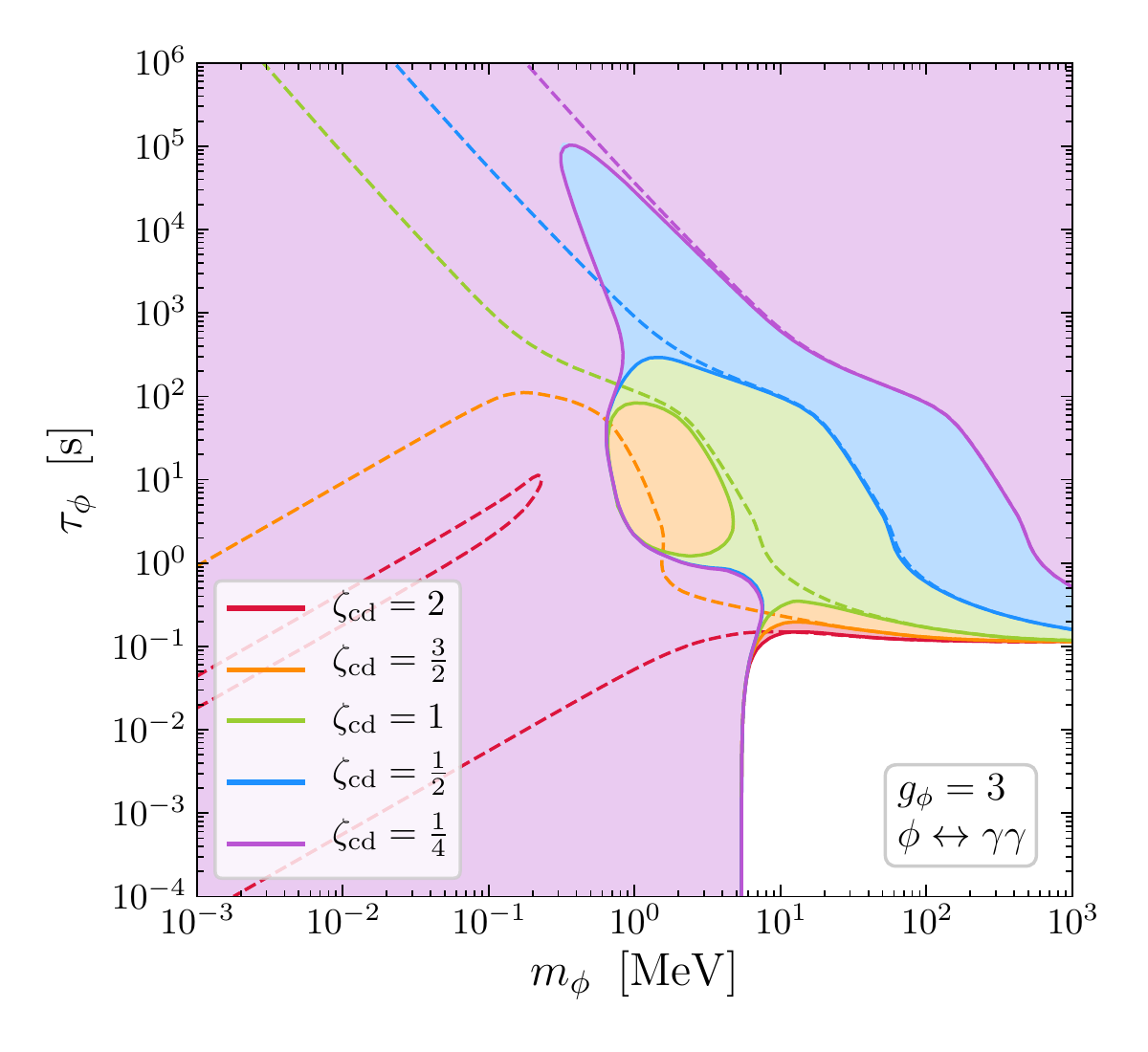}
	\includegraphics[width=0.495\textwidth]{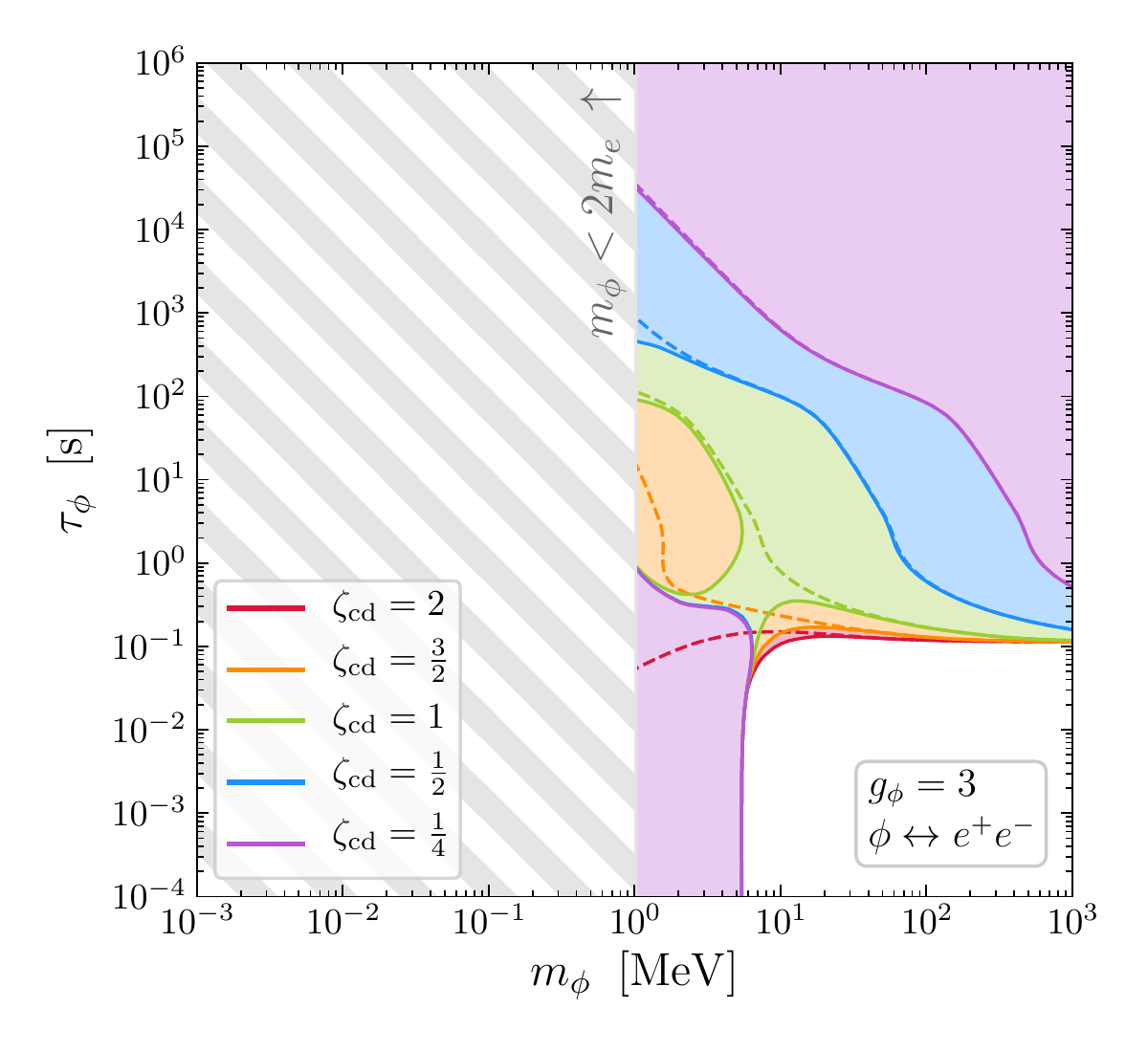}
	\caption{95\% C.L.\ constraints in $m_\phi - \tau_\phi$ plane for $g_\phi=1$, 2 and 3 degrees of freedom respectively, decaying into two photons (left) or electron-positron pairs (right) for different values of
	$\zeta_\text{cd} $ assuming $T_\text{cd} = 10 \, \mathrm{GeV}$. For comparison we also show how the limits would look when neglecting inverse decays as well as spin-statistical factors (dashed).}
\label{fig:xi_diff}
\end{figure}
\noindent $m_\phi \gtrsim 4 \, \mathrm{MeV}$ and $\tau_\phi \gtrsim 10^4 \, \mathrm{s}$ is potentially constrained.
Here we do observe a significant dependence on $T_\text{cd}$. For small values of $T_\text{cd}$ freeze-out happens when $\phi$ has already become non-relativistic and hence Boltzmann-suppressed.\footnote{Note that the relevant temperature here is the DS temperature, so that for $\zeta_\text{cd}=0.01$ Boltzmann suppression occurs correspondingly earlier.}
The abundance, in this case, is therefore predominantly determined by the `freeze-in' abundance induced by inverse decays, which is independent of $T_\text{cd}$. For $T_\text{cd} = 10 \, \mathrm{MeV}$ the exclusion region is basically identical to the case of $\zeta_\text{cd}=0$, i.e.\ the pure freeze-in scenario,  cf.\ figure~\ref{fig:freeze_in} below.
For larger values of $T_\text{cd}$, on the other hand, there is a significant `freeze-out' contribution leading to an overall larger abundance $n_\phi$, implying stronger constraints.

Let us now focus on how different values of $\zeta_\text{cd}$ impact our results.
We show the corresponding overall 95\% C.L. BBN constraint in figure~\ref{fig:xi_diff} for the same decay scenarios and $T_\text{cd} = 10 \, \mathrm{GeV}$ as in figure~\ref{fig:xi_1}, but for different values of $\zeta_\text{cd}$. The constraints arising without taking into account inverse decays and spin-statistical factors are shown as dashed lines. By direct comparison to figure~\ref{fig:nphi_ratio} it becomes evident that not taking into account these effects severely underestimates the constraints in the whole region, where inverse decays contribute appreciably to the abundance of $\phi$.
However, we also find regions, where the constraints are slightly weakened (e.g.\ for $\zeta_\text{cd}=2$ and $\phi \to \gamma \gamma$ one can observe this for $\tau_\phi \simeq 1 \, \mathrm{s}$, $m_\phi \simeq 2 \, \mathrm{MeV}$). In this case, inverse decays are largely irrelevant for the abundance $n_\phi$ and the reason for the slight weakening are thus only the spin-statistical factors for the decays leading to a faster, i.e.\ earlier, decay, cf.\ eq.~\eqref{eq:Boltz_eq_gen} with $z = \bar{z} = \gamma$ and corresponding $+$ signs.

The constraints become stronger for increasing $\zeta_\text{cd}$ due to the increased $\phi$ abundance. As before, we find only small quantitative (at $m_\phi \sim 2 m_e$) and no qualitative differences between decays into two photons or electron-positron pairs for $m_\phi > 2 m_e$. Due to the equilibration of $\phi$ with the SM whilst being relativistic, the lower limit on the mass is independent of $\zeta_\text{cd}$ (and in fact also independent of $T_\text{cd} \gtrsim 10 \, \mathrm{MeV}$) for sufficiently small lifetimes $\tau_\phi$.

\begin{figure}
	\includegraphics[width=0.495\textwidth]{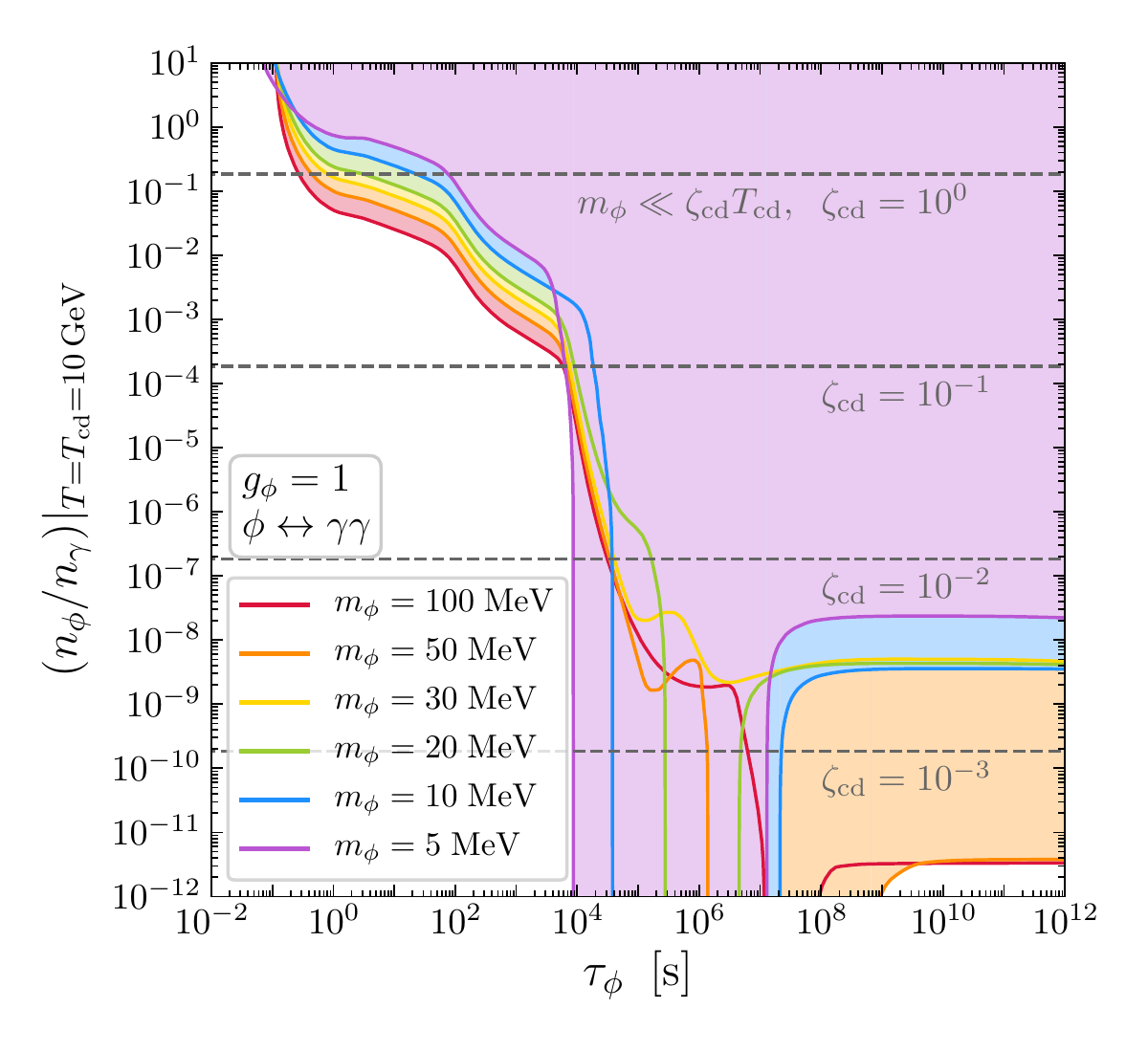}
	\includegraphics[width=0.495\textwidth]{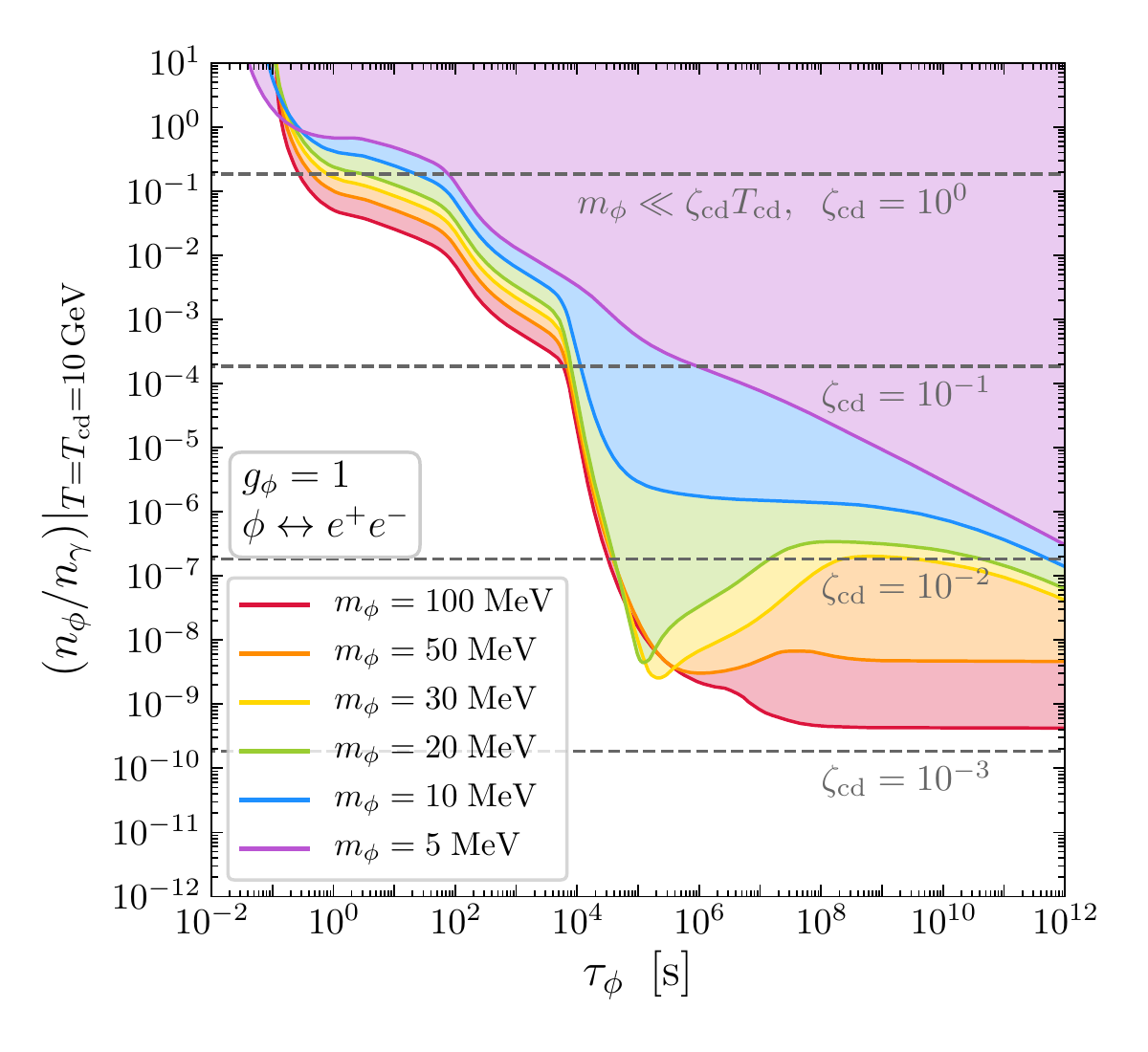}
	\caption{Overall 95\% C.L.\ BBN constraint (full) in the $\tau_\phi - (n_\phi / n_\gamma)_{T=T_\text{cd}}$ plane for a real scalar decaying into two photons (left) or an electron-positron pair (right) with different $m_\phi$ and $T_\text{cd} = 10 \, \mathrm{GeV}$. For reference, we also show lines of a constant temperature ratio $\zeta_\text{cd}$ for small masses $m_\phi$ where the relation between
$n_\phi|_{T=T_\text{cd}}$ and $\zeta_\text{cd}$  is independent of $m_\phi$ (grey, dashed).}
	\label{fig:em_mphi_multi_scalar}
\end{figure}

Let us now have a closer look at the dependence of the bounds on the number density $n_\phi$ as obtained for different values of $\zeta_\text{cd}$. If the dark sector is in kinetic equilibrium, the phase-space distribution is
of the Bose-Einstein form at chemical decoupling, so that there is a one-to-one correspondence between
$n_\phi|_{T=T_\text{cd}}$ and $\zeta_\text{cd}$ according to (cf.\ eq.~\eqref{eq:em_initial})
\begin{align}
n_\phi (T = T_\text{cd}) = g_\phi \int \frac{\d^3 p}{(2 \pi)^3} \frac{1}{\exp\left(\sqrt{p^2 + m_\phi^2} / [\zeta_\text{cd} T_\text{cd}]\right) + 1}\eqsp.
\label{eq:nphi}
\end{align}
In figure~\ref{fig:em_mphi_multi_scalar} we show the overall 95\% C.L.\  constraints from BBN including photodisintegration on a scalar decaying into two photons (left) or an electron-positron pair (right) in the $\tau_\phi - (n_\phi / n_\gamma)_{T=T_\text{cd}}$ plane for different $m_\phi$ assuming $T_\text{cd} = 10 \, \mathrm{GeV}$.\footnote{Note that, due to the presence of inverse decays, we parameterize our constraints as a function of $(n_\phi / n_\gamma)_{T=T_\text{cd}}$, which is different from the quantity $E_0 (n_\phi/n_\gamma)|_{T=T_0}$ with $T_0 < 1\,\mathrm{MeV}$, which is usually employed in the previous literature regarding photodisintegration.}
For comparison we also show lines of constant $\zeta_\text{cd}$ for $m_\phi \ll \zeta_\text{cd} \,T_\text{cd}$. For values of $m_\phi$ where $\phi$ is still relativistic during decoupling, i.e.\ for sufficiently small values of $m_\phi$, the relation between
$n_\phi|_{T=T_\text{cd}}$ and $\zeta_\text{cd}$  is independent of $m_\phi$. For larger $m_\phi$ there is a dependence on $m_\phi$, as can be seen in figure~\ref{fig:em_tau_multi_scalar} below.
Before discussing the constraints in detail, let us note that while the constraints shown implicitly assume a Bose-Einstein distribution at chemical decoupling with corresponding temperature $\zeta_\text{cd} T_\text{cd}$,  they can also to very good approximation be applied to other (non-thermal) distributions as long as deviations from this distribution are not too large, i.e.\ momenta around $\zeta_\text{cd} T_\text{cd}$ still dominate the integral in eq.~\eqref{eq:nphi}.

Let us also stress again that the number density at chemical decoupling can be substantially different from the number density at later times.
In particular, for small values of $\zeta_\text{cd}$, it is natural for $n_\phi$ to acquire a significant extra contribution shortly after chemical decoupling due to `freeze-in' from inverse decays as discussed in section~\ref{sec:evo_phi}.

The constraints start around $\tau_\phi \sim 0.1 \, \mathrm{s}$, cf.\ figure~\ref{fig:xi_diff}. For $m_\phi = 5 \, \mathrm{MeV}$ we find small quantitative differences of the different decay channels at these lifetimes. This is due to the electrons and positrons still being slightly relativistic at decay and the resulting relevance of spin-statistical factors, i.e.\ Pauli blocking. With increasing lifetime $0.1 \, \mathrm{s} \lesssim \tau_\phi < 10^4 \, \mathrm{s}$ and increasing mass $m_\phi$ the limits become stronger and are independent of the decay channel.

For $\tau_\phi \gtrsim 10^4 \, \mathrm{s}$ photodisintegration starts to become relevant.\footnote{We have neglected constraints from CMB $\mu$- and $y$-distortions, which are generically weaker~\cite{Chluba:2011hw,Chluba:2013pya,Poulin:2016anj,Chluba:2020oip} for masses above the lowest photodisintegration threshold. CMB constraints are however stronger for $\tau_\phi > 10^{12} \, \mathrm{s}$.} Depending on the mass, even the abundance produced by inverse decays only can be excluded, i.e.\ arbitrarily small values of $(n_\phi / n_\gamma)_{T=T_\text{cd}}$.
The constraints from the decay $\phi \to \gamma\gamma$ are significantly stronger compared to the limits resulting from $\phi \to e^+ e^-$, i.e.\ the resulting non-thermal photon spectrum is by far more efficient in disintegrating the nuclei in the former case.
This is especially visible as the limiting case $(n_\phi / n_\gamma)_{T=T_\text{cd}} \to 0$ is never constrained for decays into electron-positron pairs.
As photodisintegration is subject to different disintegration thresholds the constraints have a non-trivial dependence on $m_\phi$.

\begin{figure}
	\includegraphics[width=0.495\textwidth]{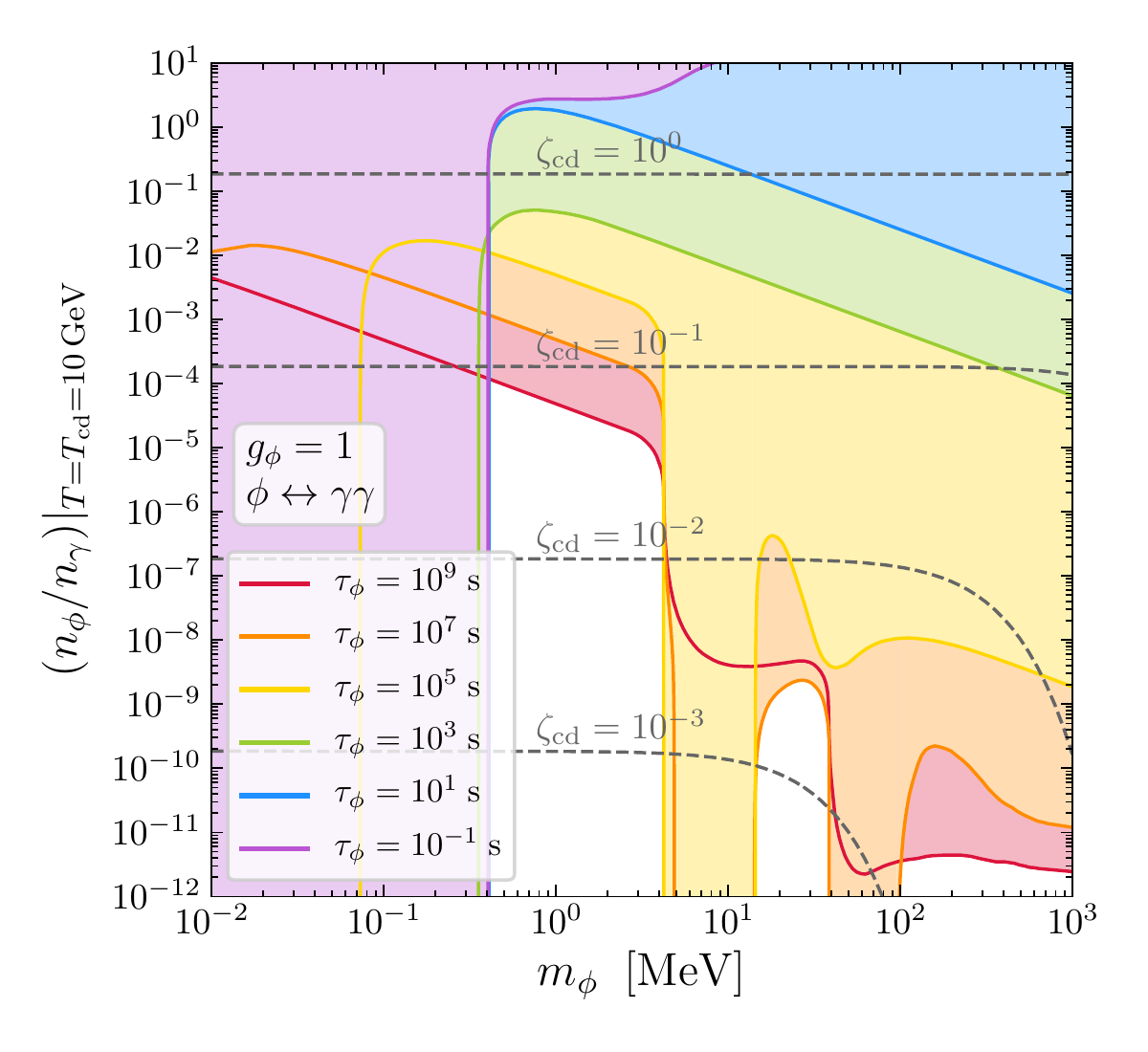}
	\includegraphics[width=0.495\textwidth]{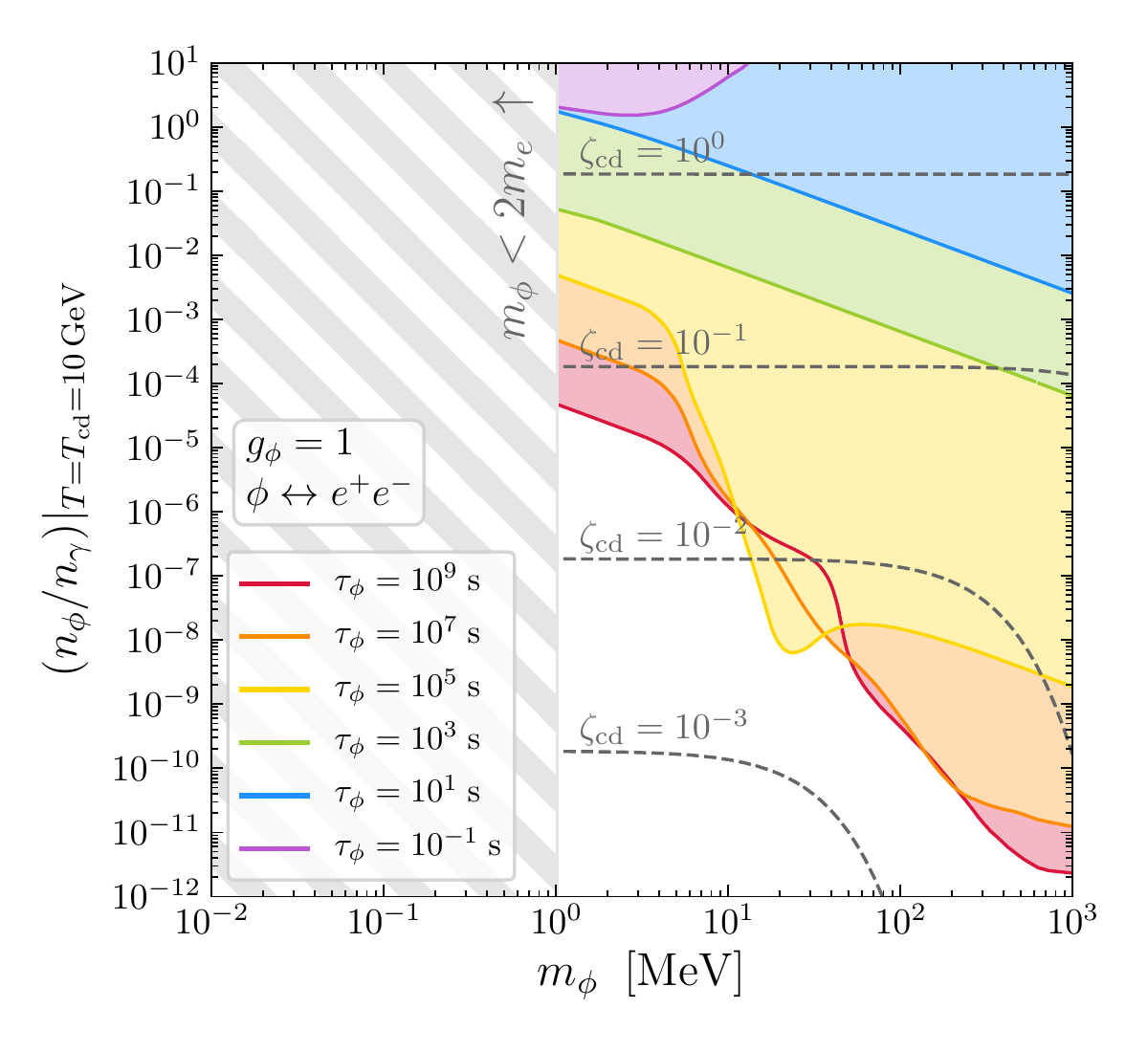}
	\caption{Overall 95\% C.L.\ BBN constraint (full) in the $m_\phi - (n_\phi / n_\gamma)_{T=T_\text{cd}}$ plane for a scalar decaying into two photons (left) or an electron-positron pair (right) with different $\tau_\phi$ and $T_\text{cd} = 10 \, \mathrm{GeV}$. For reference, we also show lines of a constant temperature ratio $\zeta_\text{cd}$ (grey, dashed).}
	\label{fig:em_tau_multi_scalar}
\end{figure}

In figure~\ref{fig:em_tau_multi_scalar} we show the overall 95\% C.L.  constraints from BBN including photodisintegration on a scalar decaying into two photons (left) or an electron-positron pair (right) in the $m_\phi - (n_\phi / n_\gamma)_{T=T_\text{cd}}$ plane for different $\tau_\phi$ assuming $T_\text{cd} = 10 \, \mathrm{GeV}$. As before, photodisintegration leads to very stringent constraints, potentially even excluding the $\phi$ abundance produced only by `freeze-in' via inverse decays corresponding to arbitrarily small $(n_\phi / n_\gamma)_{T=T_\text{cd}}$ if $\phi$ decays into photons.

As an additional cross-check of our results, we explicitly confirmed that in cases where the constraints are dominated by photodisintegration, our results resemble those from Refs.~\cite{Kawasaki:2020qxm, Forestell:2018txr} once mapped to a temperature shortly before decay. This agreement is expected as for the corresponding very long lifetimes the contribution from inverse decays is negligible.

\subsection{Constraints for freeze-in ($\zeta_\text{cd} = 0$)}

\begin{figure}[H]
	\includegraphics[width=0.495\textwidth]{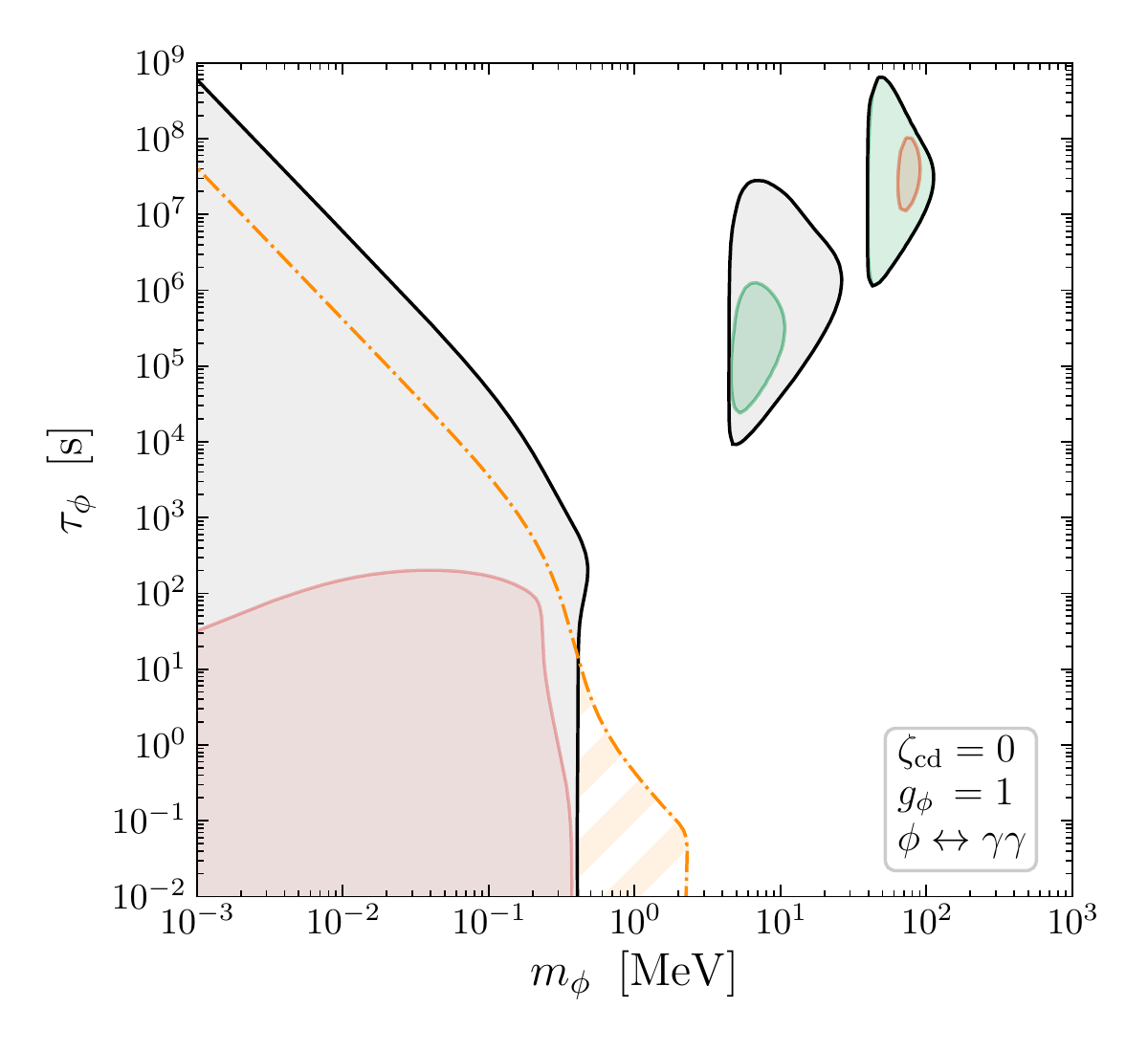}
	\includegraphics[width=0.495\textwidth]{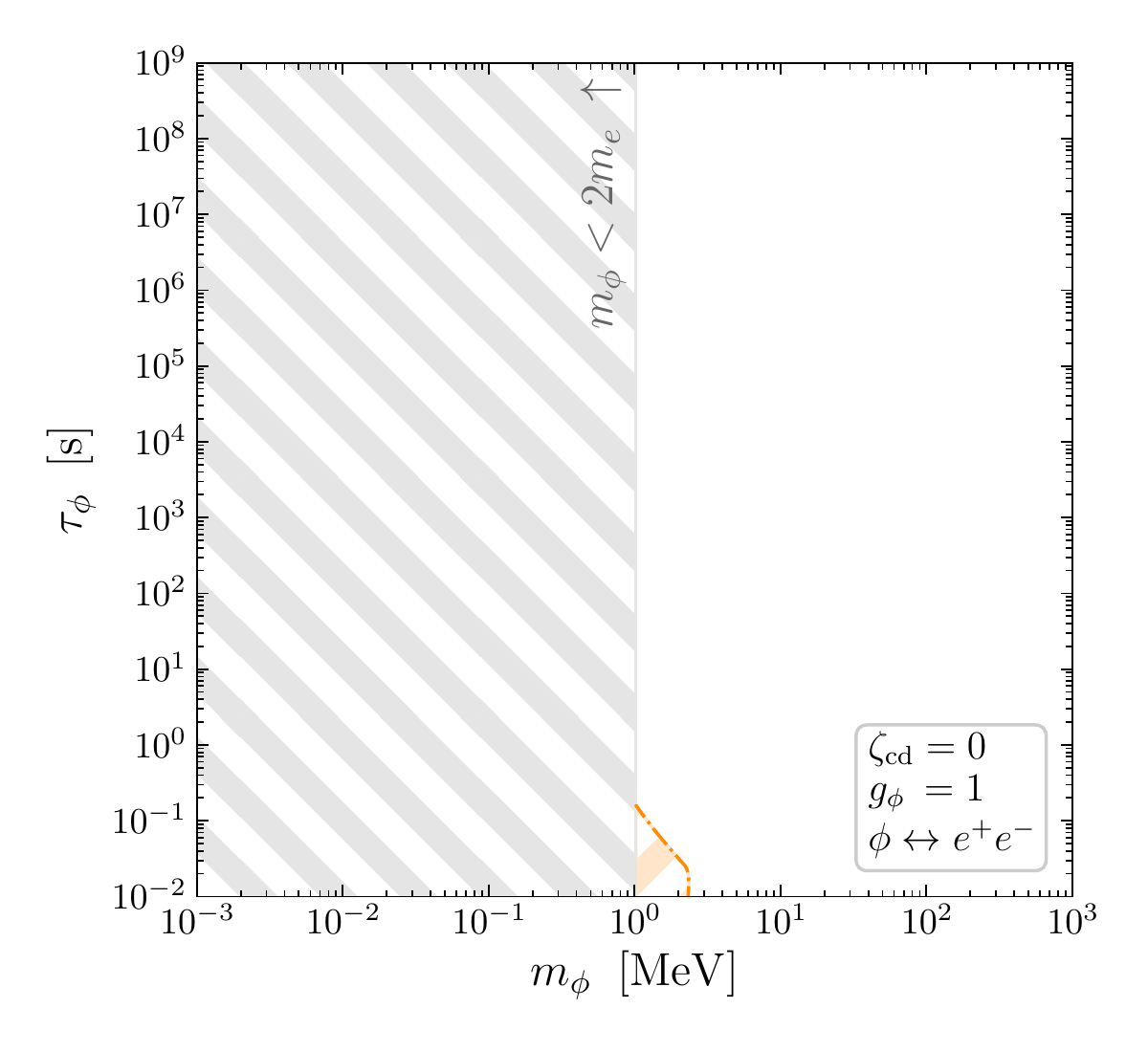}
	\includegraphics[width=0.495\textwidth]{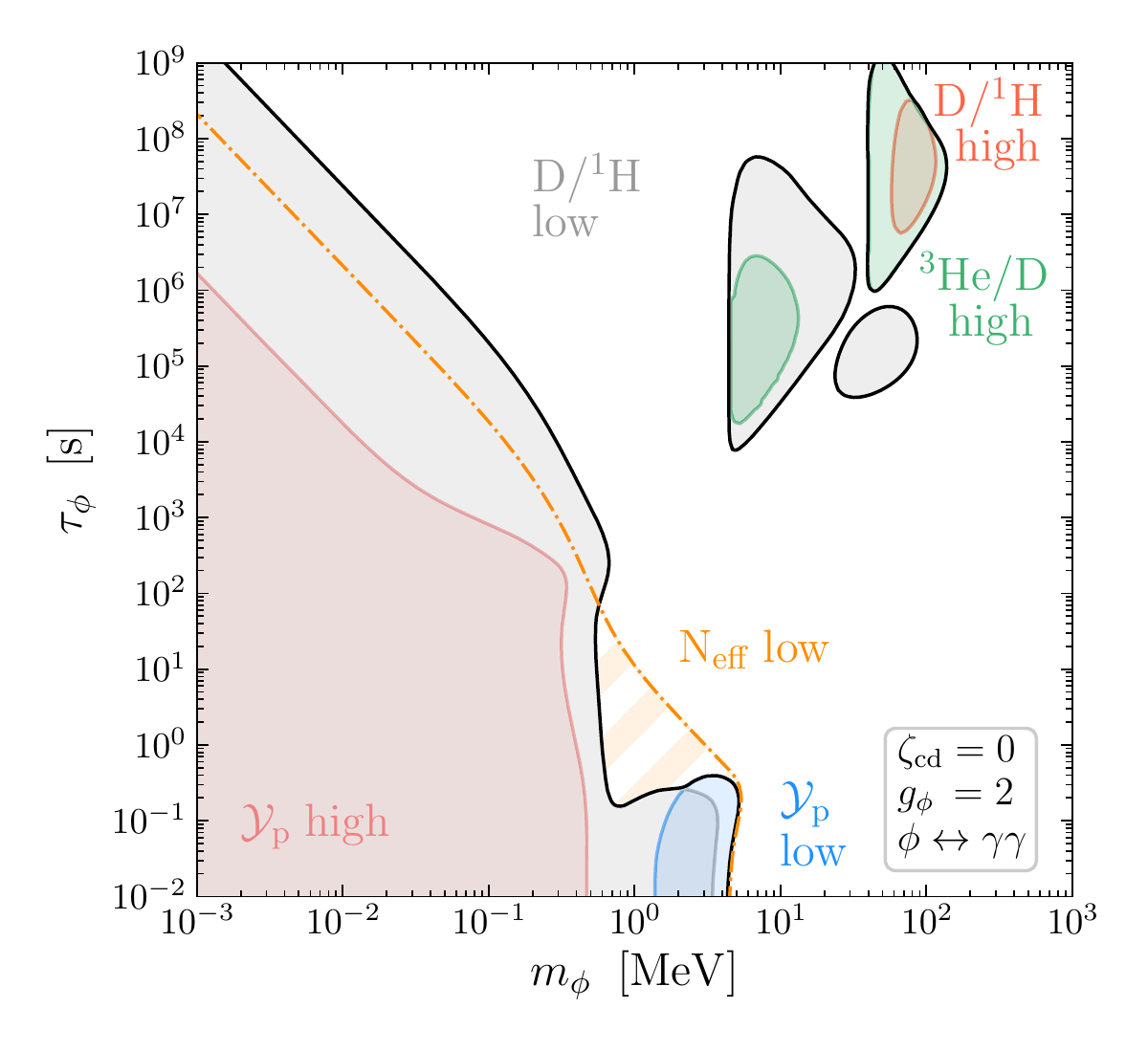}
	\includegraphics[width=0.495\textwidth]{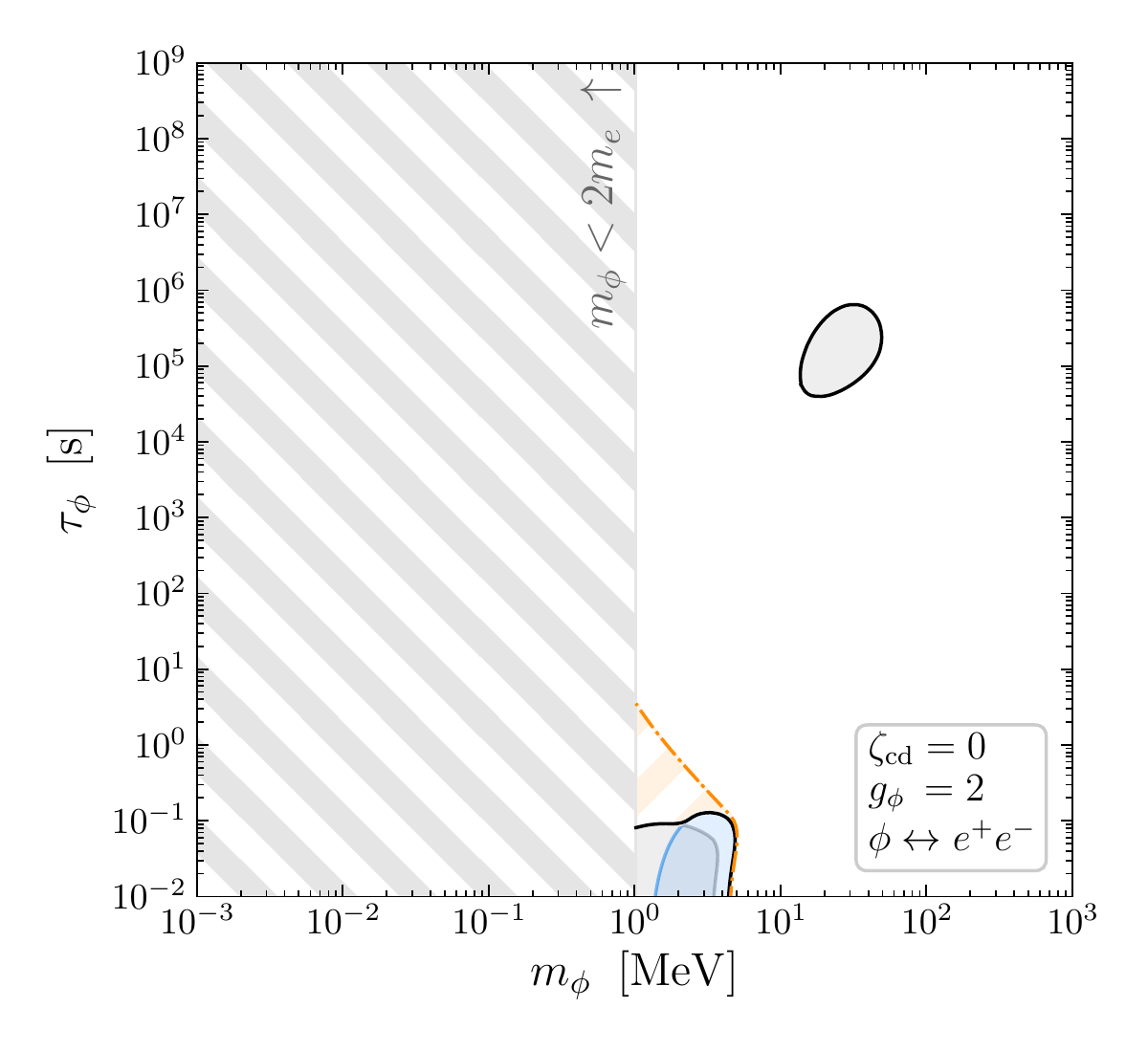}
	\includegraphics[width=0.495\textwidth]{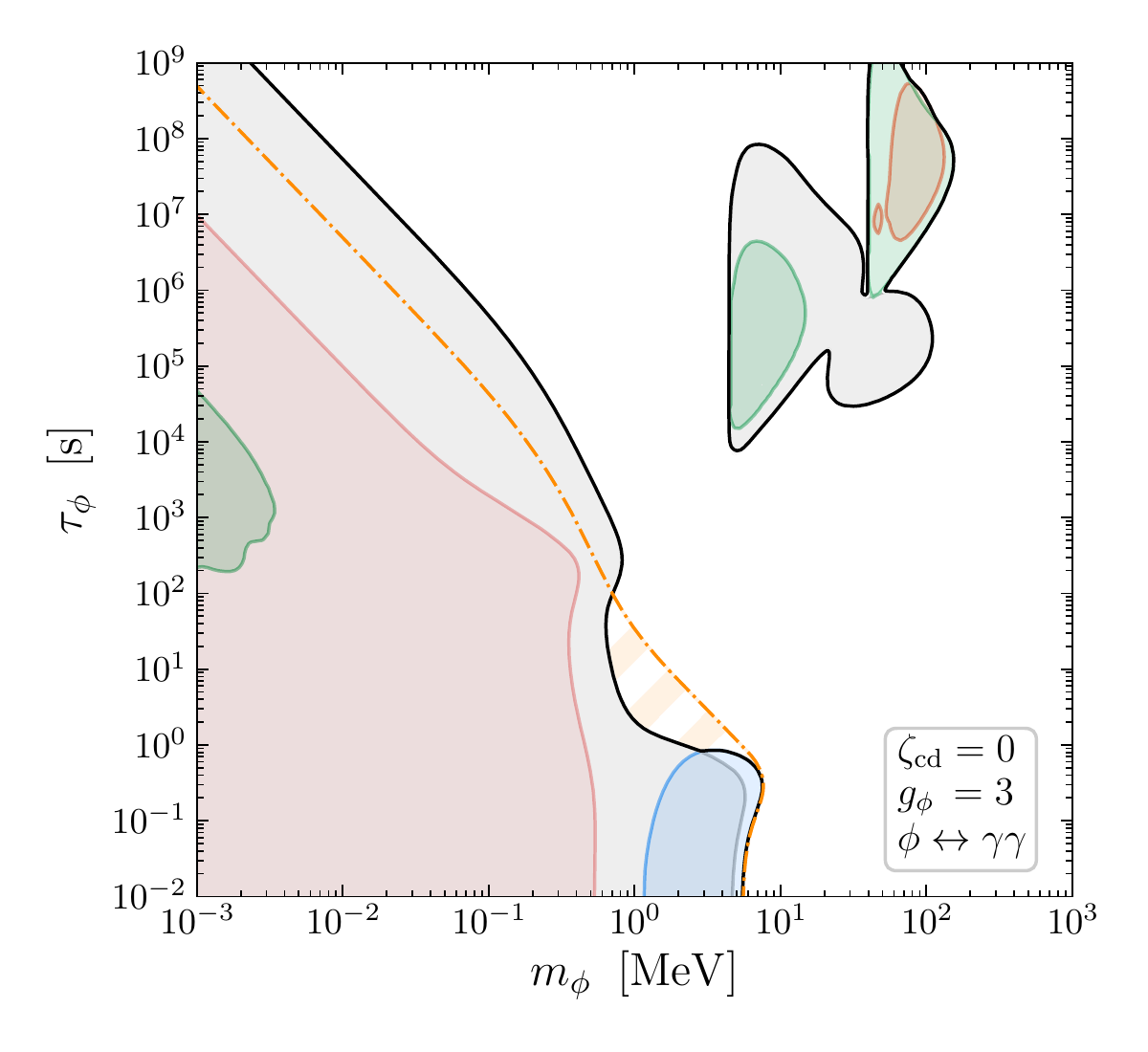}
	\includegraphics[width=0.495\textwidth]{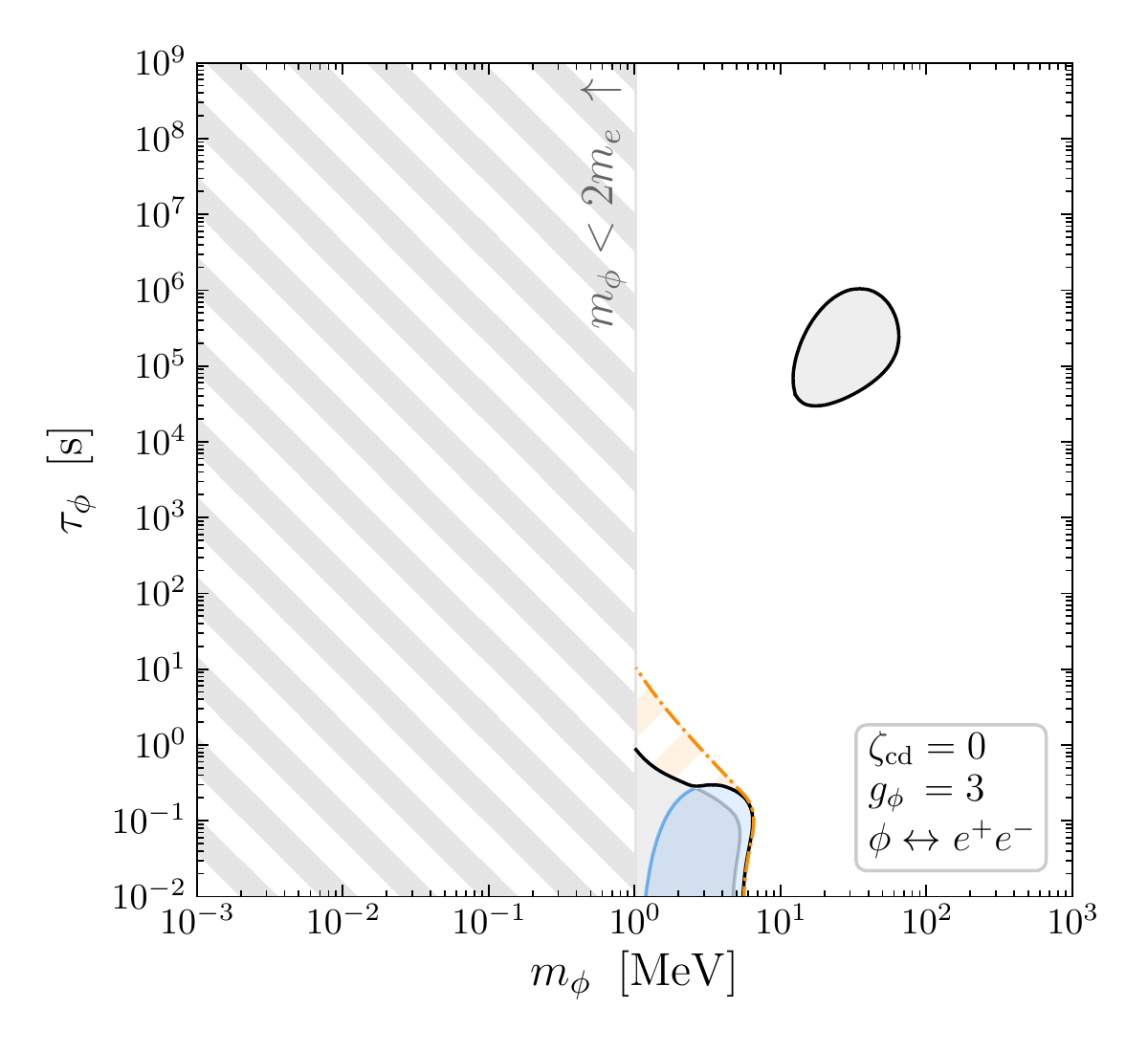}
	\caption{Same as figure~\ref{fig:xi_1} but only considering the freeze-in contribution to the $\phi$-abundance, i.e.\ $\zeta_\text{cd} = 0$.}
\label{fig:freeze_in}
\end{figure}

In figure~\ref{fig:freeze_in} we show the constraints from different primordial element abundances as well as Planck $N_\mathrm{eff}$ considering only the freeze-in contribution to the $\phi$-abundance, i.e.\ $\zeta_\text{cd} = 0$. Note that in this case photodisintegration excludes islands of parameter space for $m_\phi \gtrsim 4 \, \mathrm{MeV}$, $\tau_\phi \gtrsim 10^4 \, \mathrm{s}$ as it is sensitive to even very small abundances. The case where $\phi$ decays into photons is more strongly constrained than $\phi$ decaying into electrons due to the more energetic photons available.
For rather small lifetimes inverse decays become very effective and lead to thermalisation of $\phi$ with the SM so that the limits for small lifetimes are in fact the same as those in figure~\ref{fig:xi_1}.

\section{Implications for the Li problem}
\label{sec:li}

As mentioned above there is a long-standing discrepancy between the SM prediction of the amount of primordial lithium and the value inferred from astrophysical observations~\cite{Fields:2011zzb}. The origin of this discrepancy is still unresolved, but as large astrophysical uncertainties may prohibit an unbiased measurement of the {\it primordial} lithium abundance, we conservatively only took into account the limits from hydrogen, deuterium and $^3$He in the preceding sections.

In this section, we will instead consider the possibility that this discrepancy, in fact, points to some new physics which changes the theoretical prediction of the lithium abundance.
Specifically, we will concentrate on decaying $\mathrm{MeV}$-scale particles, making use of the fact that the photodisintegration threshold for beryllium (which later converts into lithium-7) is below the one
for deuterium, implying that the beryllium abundance can be significantly reduced while leaving the other element abundances practically unchanged~\cite{Poulin:2015woa,Salvati:2016jng,Kawasaki:2020qxm}.
More quantitatively, the photon energy threshold for the reaction ${}^7\text{Be}  \gamma \rightarrow {}^3\text{He} {}^4\text{He}$ is $1.59 \, \mathrm{MeV}$ while the photon energy threshold for the reaction $D \gamma \rightarrow n p$ is $2.22 \, \mathrm{MeV}$. As a large fraction of the cosmological ${}^7\text{Li}$ abundance originates from production and subsequent decay of ${}^7\text{Be}$ this opens up the possibility of bringing down the ${}^7\text{Li}$ abundance via photodisintegration of ${}^7\text{Be}$. This may solve the Li problem without underproducing D if $3.17 \, \mathrm{MeV} < m_\phi < 4.44 \, \mathrm{MeV}$.  As soon as the mass becomes larger than $m_\phi > 4.44 \, \mathrm{MeV}$, photodisintegration of deuterium typically excludes the region which would be favourable for lithium, as we will see below. In the following we will concentrate on the decay $\phi \to \gamma \gamma$, as decays into electron-positron pairs lead to less efficient photodisintegration and we checked that in this case no viable parameter space remains.

\begin{figure}
	\includegraphics[width=0.495\textwidth]{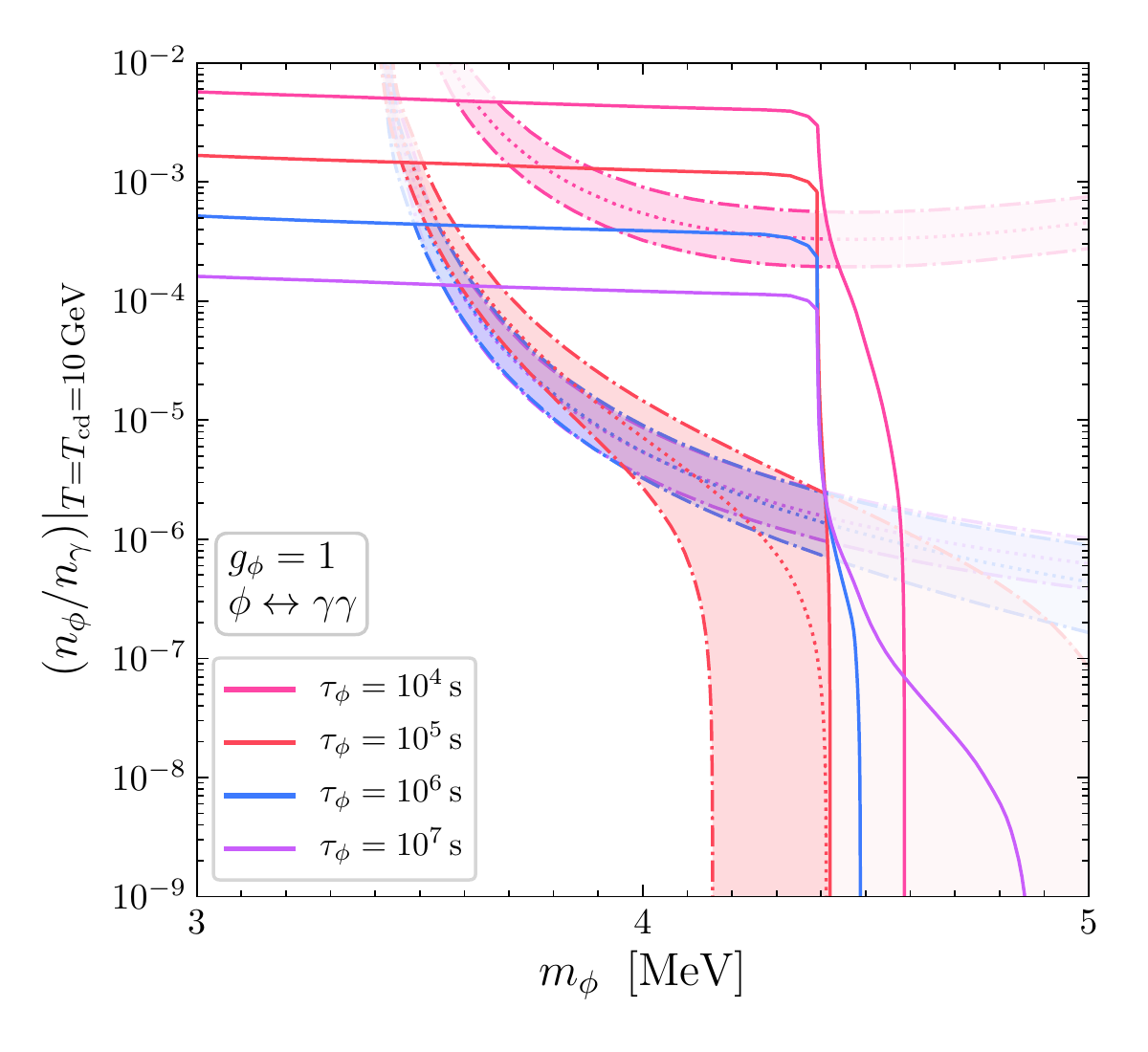}
	\includegraphics[width=0.495\textwidth]{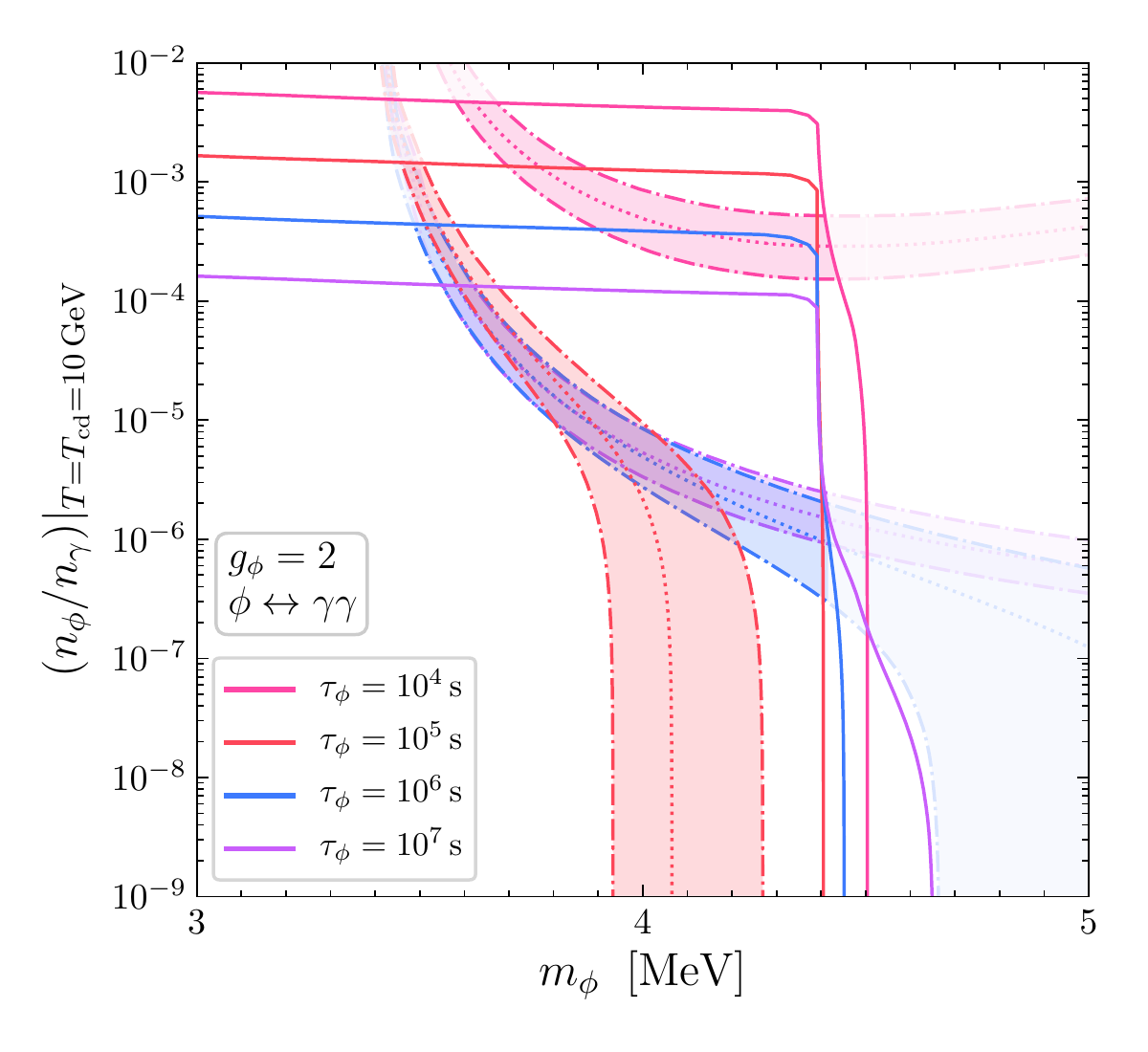}
	\caption{Combined 95\% C.L. BBN constraint from D, ${}^3\text{He}$ and ${}^4\text{He}$ (full line) and region where the Li problem is solved (filled regions -- for the dotted line the central value of the observation is reproduced, the dash-dotted lines correspond to 95\% C.L.) in the $m_\phi - (n_\phi / n_\gamma)_{T=T_\text{cd}}$ plane for decays
	into two photons and $g_\phi=1$ (left) and $g_\phi=2$ (right) assuming $T_\text{cd} = 10 \, \mathrm{GeV}$.}
\label{fig:li_scalar}
\end{figure}

\begin{figure}
	\includegraphics[width=0.495\textwidth]{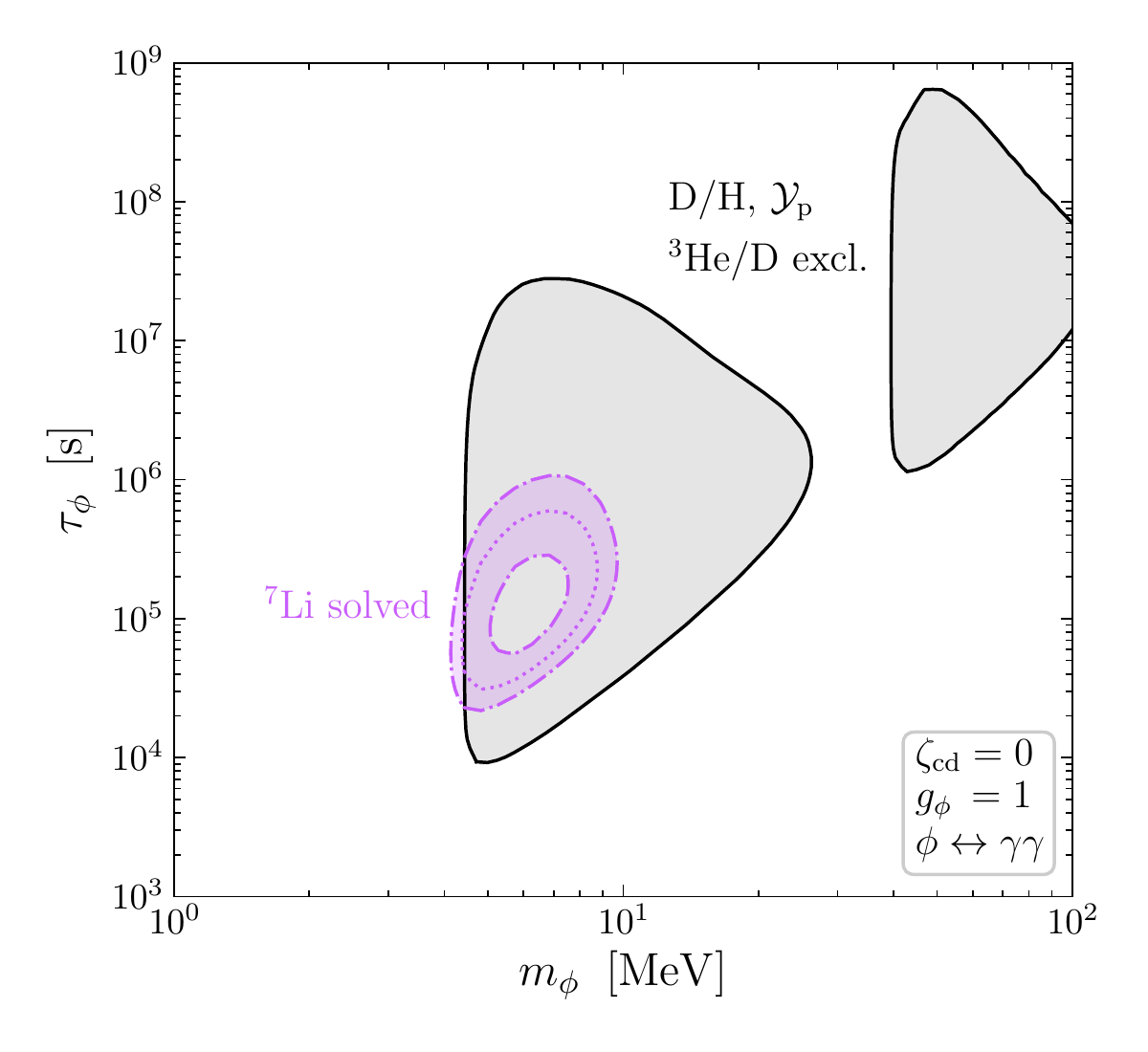}
	\includegraphics[width=0.495\textwidth]{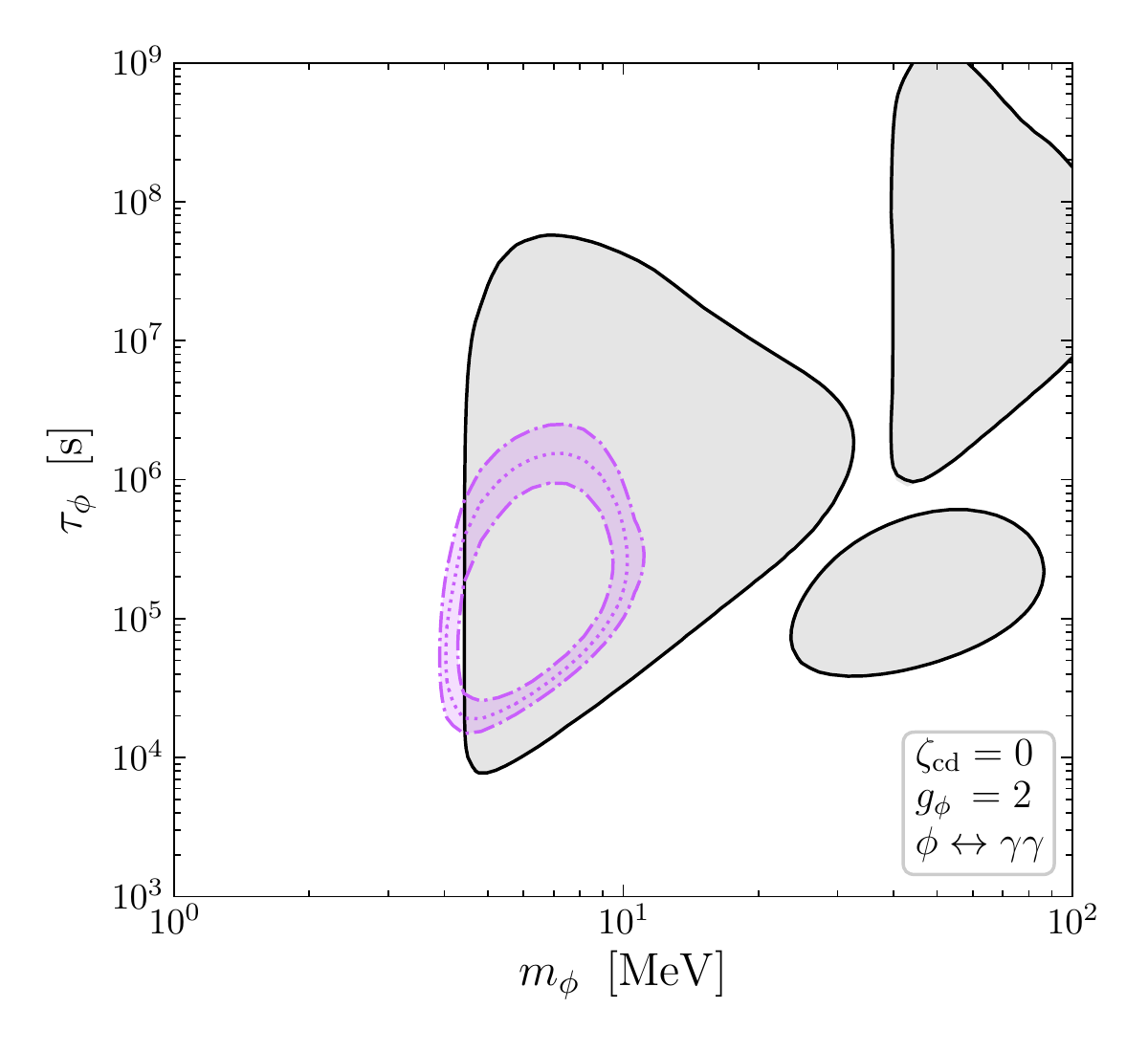}
	\caption{Combined 95\% C.L. BBN constraint from D, ${}^3\text{He}$ and ${}^4\text{He}$ (full black line) and region where the Li problem is solved (filled violet regions)
	in the $m_\phi - \tau_\phi$ plane for decays into two photons and $g_\phi=1$ (left) and $g_\phi=2$ (right) assuming $\zeta_\text{cd} = 0$, i.e.\  a pure freeze-in contribution.}
	\label{fig:li_freeze_in}
\end{figure}

\begin{figure}
	\includegraphics[width=0.495\textwidth]{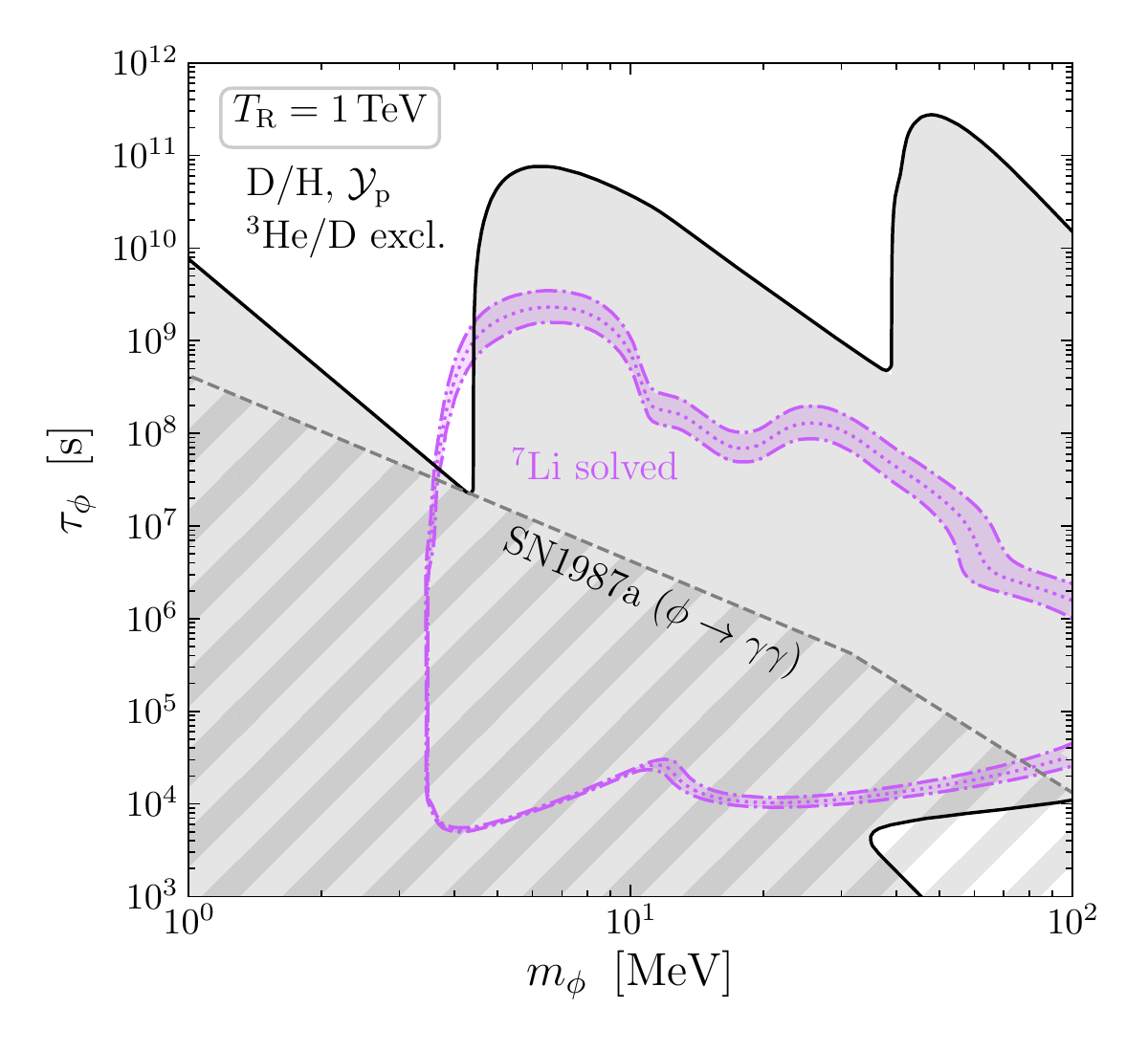}
	\includegraphics[width=0.495\textwidth]{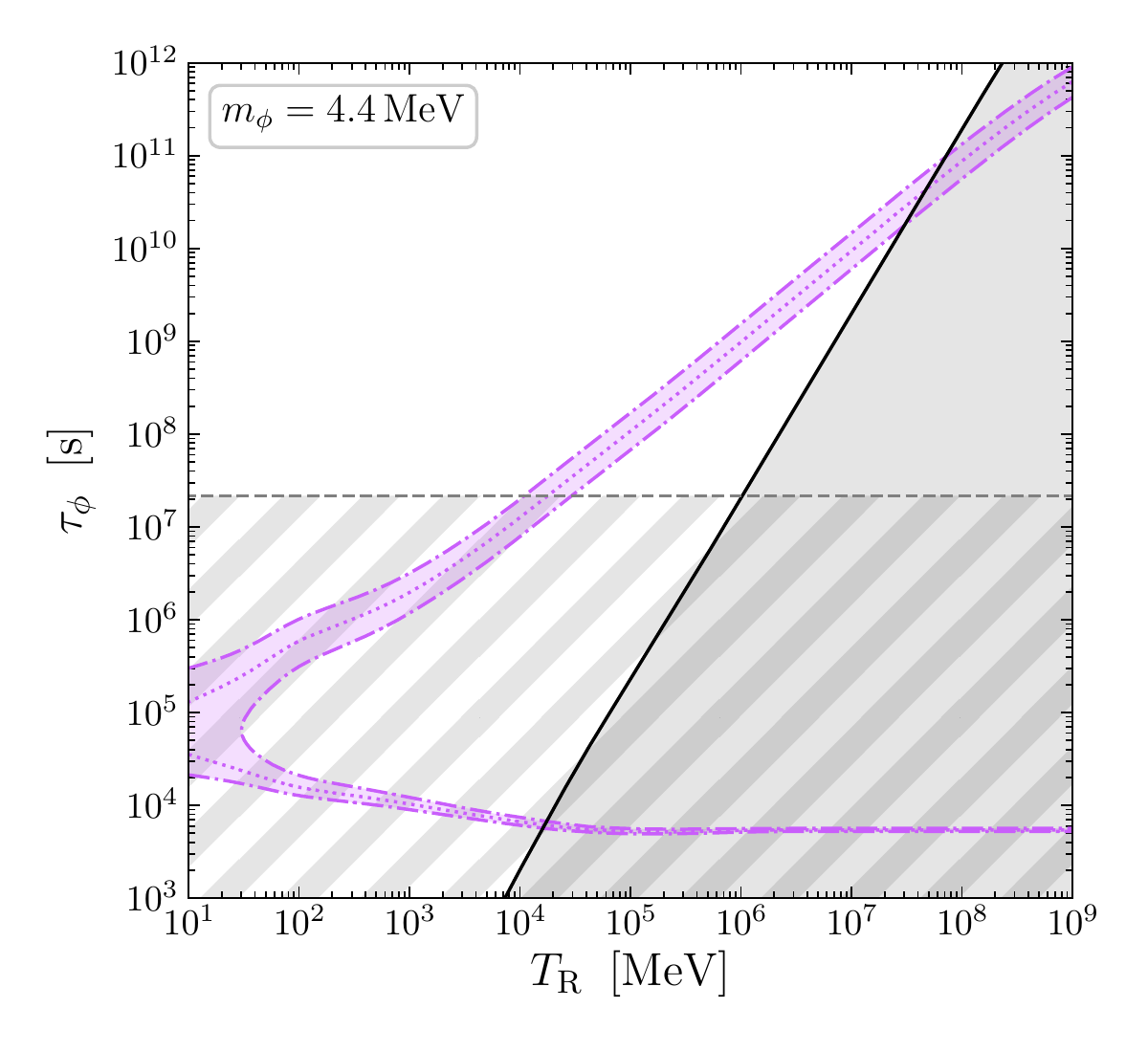}
	\caption{Combined 95\% C.L. BBN constraint from D, ${}^3\text{He}$ and ${}^4\text{He}$ (full black line) and region where the Li problem is solved (filled violet
	regions) for an ALP decaying into two photons. In the left panel we fix the reheating temperature to $T_\mathrm{R} = 1 \, \mathrm{TeV}$ while in the right panel we fix $m_\phi = 4.4 \, \mathrm{MeV}$.
	For comparison we also show constraints from visible decays of ALPs produced in the supernova SN1987a (hatched region).}
	\label{fig:li_alp}
\end{figure}

While~\cite{Poulin:2015woa,Kawasaki:2020qxm} concentrate on a heuristic assessment (simply assuming a freely adjustable comovingly constant particle number density before decay),
we concentrate on a consistent evolution of the number density in a number of different setups.
Specifically, we consider the following different possibilities for the decaying particle $\phi$:
\begin{enumerate}
\item[{\it i)}] a decaying particle initially in equilibrium with the dark sector (with a temperature different from the SM), cf.\ figure~\ref{fig:li_scalar},
\item[{\it ii)}] a decaying particle produced only via the model-independent freeze-in interactions due to inverse decays, cf.\ figure~\ref{fig:li_freeze_in}, and
\item[{\it iii)}] an ALP produced via the Primakoff interaction and subsequent decay into two photons, cf.\ figure~\ref{fig:li_alp}.
\end{enumerate}

{\it i) \; --- \; } Let us start with the case where $\phi$ is in thermal equilibrium with a dark sector, see figure~\ref{fig:li_scalar}.  We show the combined 95\% C.L. BBN constraint from D, ${}^3\text{He}$ and ${}^4\text{He}$ as a full line for different lifetimes $\tau_\phi$ as a function of the mass $m_\phi$ and the number density $n_\phi$ at the decoupling temperature $T_\text{cd} = 10 \, \mathrm{GeV}$.
In the coloured region bounded by a dash-dotted line the predicted lithium-7 abundance is in agreement with observation at 95\% C.L.\ (for the dotted line the central value of the observation is reproduced). The part of parameter space in which this is also in agreement with the abundances of
D, ${}^3\text{He}$ and ${}^4\text{He}$ is further highlighted and corresponds to the region where the lithium problem is fully resolved. We observe that for the chosen $T_\text{cd} = 10 \, \mathrm{GeV}$
the abundance of $\phi$ corresponds to a DS temperature which has to be smaller than the one in the SM. In fact even very suppressed number densities at $T_\text{cd} = 10 \, \mathrm{GeV}$
still allow for a solution of the lithium problem due to an irreducible freeze-in contribution induced by inverse decays.

{\it ii)  \; --- \; } Assuming that the abundance of $\phi$ was dominantly produced via freeze-in removes the dependence on the initial temperature (or number density) as a parameter such that for a given particle type and decay channel we are only left with the mass and lifetime as free parameters. Interestingly, as can be seen in figure~\ref{fig:li_freeze_in}, the lithium problem can also be solved in this more restricted setup without being in conflict with other primordial element abundance observations if the mass is just between twice the disintegration thresholds of ${}^7\text{Be}$ and D, $3.17 \, \mathrm{MeV} < m_\phi < 4.44 \, \mathrm{MeV}$. Again this is only possible for decays into photons and not into electrons as discussed before.  As the abundance of $\phi$ is significantly below the thermal abundance, the scenario can only be constrained via photodisintegration leading to characteristic exclusion `islands' starting at twice the threshold energies.

{\it iii)  \; --- \; } Let us finally consider the case of an axion-like particle (ALP) which couples predominantly to photons. We follow the procedure detailed in~\cite{Depta:2020wmr} for the calculation of the cosmological evolution. The ALP abundance is determined both by Primakoff interactions, $q^\pm \phi \leftrightharpoons q^\pm \gamma$ (with $q^\pm$ a charged SM particle), and inverse decays.
The Primakoff interactions will establish thermal equilibrium with the SM for sufficiently large reheating temperatures $T_\mathrm{R}$.
However, in the relevant mass region $3.17 \, \mathrm{MeV} < m_\phi < 4.44 \, \mathrm{MeV}$ and for lifetimes with relevant contributions to photodisintegration $\tau_\phi \gtrsim 10^4 \, \mathrm{s}$, the parameter space of ALPs coupled predominantly to photons is excluded as long as the reheating temperature $T_\mathrm{R}$ after cosmic inflation is above the freeze-out temperature of the Primakoff interaction, see figure 4 of~\cite{Depta:2020wmr}. The lithium problem in this setup can hence only be solved for a smaller $T_\mathrm{R}$ such that the dominant production is a `freeze-in' contribution via inverse decays. In the left panel of figure~\ref{fig:li_alp} we choose an exemplary value of $T_\mathrm{R} = 1 \, \mathrm{TeV}$ and show the overall 95 \% C.L.\ BBN constraints and the region, where the lithium problem can be solved, in the $m_\phi$-$\tau_\phi$ plane. We also show the constraints from visible decays of ALPs produced in the supernova SN1987a~\cite{Jaeckel:2017tud} (dashed). To stress that their reliability is currently under debate~\cite{Bar:2019ifz} we use a hatched filling. For small mass and lifetime, the ALP is still efficiently produced via Primakoff interactions and (inverse) decays and thus excluded by BBN.
For larger masses and lifetimes we again observe characteristic exclusion `islands' starting at twice the threshold energies, but also regions where the lithium problem can be solved without being in conflict with other primordial element abundance observations.

In the right panel we fix the ALP mass to $m_\phi = 4.4 \, \mathrm{MeV}$, just below the maximal viable mass below the D disintegration threshold, and vary the reheating temperature and the ALP lifetime. BBN excludes a region towards large $T_\mathrm{R}$ and small $\tau_\phi$ as this leads to efficient ALP production via the Primakoff process while visible decays of ALPs produced in the SN1987a exclude $\tau_\phi \lesssim 2.2 \times 10^7 \, \mathrm{s}$ for this mass.
The lithium problem can be solved for reheating temperatures as low as the lower limit from BBN, $T_R \gtrsim 10 \, \mathrm{MeV}$~\cite{Hasegawa:2019jsa} and up to $T_\mathrm{R} \sim 5 \times 10^7 \, \mathrm{MeV}$.
The allowed lifetimes lie in the range $6 \times 10^3 \, \mathrm{s} \lesssim \tau_\phi \lesssim 7 \times 10^{10} \, \mathrm{s}$. Taking the SN1987a constraints at face value restricts the lower limits to $T_\mathrm{R} \sim 10^4 \, \mathrm{MeV}$ and $\tau_\phi \sim 2.2 \times 10^7 \, \mathrm{s}$, respectively.

\section{Conclusions}
\label{sec:conclusions}

In this work, we revise and update model-independent constraints from Big Bang Nucleosynthesis and photodisintegration processes on MeV-scale particles $\phi$ which decay into photons and/or electron-positron pairs, $\phi \to \gamma\gamma$ and $\phi \to e^+ e^-$. In parallel to this article, we also release the public code \texttt{ACROPOLIS} which numerically solves the reaction network necessary to evaluate the effect of photodisintegration on the final light element abundances.
In the current study, we pay particular attention to a self-consistent cosmological evolution of the phase-space distribution of $\phi$, including {\it model-independent} contributions to the $\phi$ abundance $n_\phi$ from inverse decays.
We also include all spin-statistical factors with full Bose-Einstein and Fermi-Dirac distribution functions.
We find that taking these issues as well as the latest determinations of primordial abundances into account, the bounds become significantly stronger compared to \cite{Hufnagel:2018bjp} in large regions of parameter space.
For a dark sector which has a similar temperature as the SM sector at chemical decoupling of $\phi$, i.e.\ a temperature ratio $\zeta_\text{cd}\sim1$, the bounds significantly strengthen in particular for small masses.
For the case of a much colder dark sector, $\zeta_\text{cd} \ll 1$, bounds become stronger almost everywhere.
In fact we find that even for $\zeta_\text{cd} =0$ (i.e.\ a pure freeze-in abundance due to inverse decays) significant limits remain.
Let us also stress that inverse decays often non-trivially change the abundance of $\phi$ {\it during} the time of BBN, and the resulting limits can therefore not be recovered assuming a comovingly
constant number density $n_\phi$ and subsequent decays with a lifetime $\tau_\phi$, as mostly used for existing BBN constraints.
We provide a large number of plots in order to allow the approximate inference of BBN limits also for other scenarios.

We also re-evaluate a possible solution of the lithium problem due to photodisintegration of beryllium, which effectively leads to a depletion of the primordial lithium abundance. As long as the final-state photons originating from the $\phi$ decay have energies below the deuterium threshold, the other abundances are largely unaffected and complete agreement with the measured primordial abundances can be achieved
for $3.17 \, \mathrm{MeV} < m_\phi < 4.44 \, \mathrm{MeV}$.
Again we concentrate on a consistent evolution of the number density in a number of different setups and find that for a low dark sector temperature,  $\zeta_\text{cd} \ll 1$, a real or complex scalar decaying into photons $\phi \to \gamma\gamma$ can lead to a viable solution. Another promising possibility is an axion-like particle coupled predominantly to the electromagnetic field strength tensor as long as the reheating temperature $T_\text{R}$ after inflation does not significantly exceed 1~TeV.

\acknowledgments

This work is supported by the ERC Starting Grant `NewAve' (638528), the
Deutsche Forschungsgemeinschaft under Germany's Excellence Strategy -- EXC 2121 `Quantum Universe' -- 390833306, and by the F.R.S. -- FNRS under the Excellence of Science (EoS) project No. 30820817 -- be.h `The H boson gateway to physics beyond the Standard Model'.

\bibliography{refs}
\bibliographystyle{JHEP}

\end{document}